\documentclass[preprint,showpacs,showkeys,aps]{revtex4}
\usepackage[english]{babel}

\usepackage[dvips]{graphicx}
\usepackage{amssymb}

\begin{document}


\title{Simulation of Two- and Three-Dimensional Dense-Fluid Shear
Flows {\em via} Nonequilibrium Molecular Dynamics.  Comparison of
Time-and-Space-Averaged Stresses from Homogeneous Doll's and Sllod
Shear Algorithms with those from Boundary-Driven Shear.}

\author{Wm. G. Hoover and Carol G. Hoover \\
Ruby Valley Research Institute \\ Highway Contract 60,
Box 598, Ruby Valley 89833, NV USA \\
Janka Petravic \\
Complex Systems in Biology Group \\
Centre for Vascular Research \\
The University of New South Wales \\
Sydney NSW 2052, Australia      }
\date{\today}




\vskip 0.5cm

\begin{abstract}

{\em Homogeneous} shear flows (with constant strainrate $dv_x/dy$) are
generated with the Doll's and Sllod algorithms and compared to
corresponding {\em inhomogeneous} boundary-driven flows.  We
use one-, two-, and three-dimensional smooth-particle weight functions
for computing instantaneous spatial averages.  The nonlinear normal
stress differences
are small, but significant, in both two and three space dimensions.
In homogeneous systems the sign and magnitude of the shearplane
stress difference, $P_{xx} - P_{yy}$, depend on both the thermostat
type and the chosen shearflow algorithm.  The Doll's and Sllod
algorithms predict opposite signs for this normal stress difference, with
the Sllod approach definitely wrong, but somewhat closer to the
(boundary-driven) truth.  {\em Neither} of the homogeneous shear
algorithms predicts the correct ordering of the kinetic temperatures:
$T_{xx} > T_{zz} > T_{yy}$.

\end{abstract}

\maketitle

\section{Introduction}

In the present work, we use nonequilibrium molecular dynamics\cite{b1}
to study microscopic simulations of ``simple shear flow'' (also called
``plane Couette flow''):
$$
v_x \propto y \rightarrow  P_{xy} \ \equiv \ P_{yx} \ \equiv \
-\eta [(\partial v_x/\partial y) + (\partial v_y/\partial x)] \
= \ -\eta (dv_x/dy) \ = \ -\eta \dot \epsilon \ .
$$
The flow is in the $x$ direction so that the tensor $\nabla v$ has only
one nonzero element, $(\nabla v)_{yx} = dv_x/dy = \dot \epsilon $.
We use the symbol $P$ for the (symmetric second-rank Cauchy) pressure
tensor (positive in compression and the negative of the stress tensor,
$P = -\sigma $); $v$ for the (time- and space-dependent)
hydrodynamic flow velocity; $\eta $ for the shear viscosity;
and $\dot \epsilon $ for the magnitude of the imposed (or measured) shear
strain rate.  The various simulation types we consider here were designed
to clarify the
relationships between periodic homogeneous methods, thermostated
{\em everywhere}, and flows with moving thermostated {\em boundaries}.
{\em All} the methods we use are consistent with Green and Kubo's linear
response theory at small rates of shear\cite{b2}.  We are specially
interested in characterizing and understanding the  {\em nonlinear}
shearplane stress difference:
[ $P_{xx} - P_{yy} = \sigma_{yy} - \sigma_{xx}$ ]
which arises in sufficiently small systems at sufficiently large
shear rates.  In carrying out microscopic simulations both the boundary
conditions and the thermostats or ergostats which control the flow need to
be carefully considered\cite{b1}.

Though simple shear flow is ``stationary'', {\em fluctuations} in local
properties necessitate {\em averaging}, both in time and in space.  Here
we reduce the importance of these fluctuations by using
spatial averaging techniques borrowed from smooth-particle continuum simulation
methods\cite{b3}.  We measure instantaneous {\em spatially-averaged} flow
velocity, temperature, and pressure-tensor components.

The two best-known homogeneous microscopic methods, the Doll's
tensor\cite{b4} and Sllod algorithms\cite{b5}, treat fluid rotation
differently, leading to qualitatively different predictions for the
nonlinear normal stress difference. {\em Boundary-driven} flows can help to
resolve this disagreement\cite{b6,b7,b8,b9}.  {\em Useful} flows
need to satisfy four conditions:
the spatial scale $L$ of these flows needs to be large enough (relative to the
particle size), but not too large (to avoid turbulence), with flow
velocities $v$ large enough (to emerge above fluctuations), but not too large
(again, to avoid turbulence), in order to provide useful information. The
relatively greater importance of fluctuations in two dimensions is
responsible for the reduced utility of viscosity there, as shown in Fig. 1.
Additionally, the frictional boundary-fluid interaction needs to be
sufficiently strong to prevent excessive boundary slip.

In the present work, we carry out both two- and three-dimensional simulations
using all three approaches (Doll's, Sllod, and boundary-driven) for two
simple, and rather similar, pairwise-additive repulsive potentials.  We
focus here on the normal
stress difference, $P_{xx} - P_{yy} = \sigma _{yy} - \sigma _{xx}$, in
the shearplane.  We will see that the Doll's and Sllod algorithms typically
predict different signs for the simple monatomic dense fluids considered
here.  Both these {\em homogeneous} algorithms yield clearcut results,
insensitive to system size.

In the early days of such shearflow simulations\cite{b10} it was thought
that the two-dimensional viscosity might depend logarithmically on system
size\cite{b11}.  More recently, careful studies\cite{b12,b13,b14,b15} of the
corresponding Green-Kubo stress-stress correlation function:
$$
\eta =
(V/kT)\int_0^\infty \langle \ P_{xy}(0)P_{xy}(t) \ \rangle_{\rm eq}dt \ ,
$$
indicate no such dependence in dense fluids.  The work we carry out here
is consistent with this lack of size dependence.  Here we find that the
coefficient of the hypothetical logarithmic viscosity contribution can
be no larger than $10^{-4}$ in the natural reduced units of atomic size,
mass, and velocity.  The more realistic {\em boundary-driven} flows
necessarily entail larger fluctuations and considerable size dependence.
We are nevertheless able to determine the sign and the size of the
normal stress difference for such flows so as to characterize the errors
inherent in the homogeneous algorithms.

This paper is organized as follows: in Sec. II we lay out the macroscopic
description of the problem; in Sec. III the {\em microscopic} description;
In Sec. IV we describe the two homogeneous
algorithms emphasizing their similarities and differences; in Sec. V we
describe the boundary-driven algorithm used in the present work; Sec. VI
describes the smooth-particle spatial averaging method used in analyzing
results from simulations;  Sec. VII describes numerical results for two
similar (smooth repulsive) forcelaw models in two very-different density
regimes and in two space dimensions; Sec. VIII describes corresponding
three-dimensional results for one of these forcelaw models;
Sec. IX lists the conclusions we have drawn from this work.

\section{Macroscopic Description of Simple Shear Flow}

\begin{figure}
\includegraphics[height=6cm,width=6cm,angle=-90]{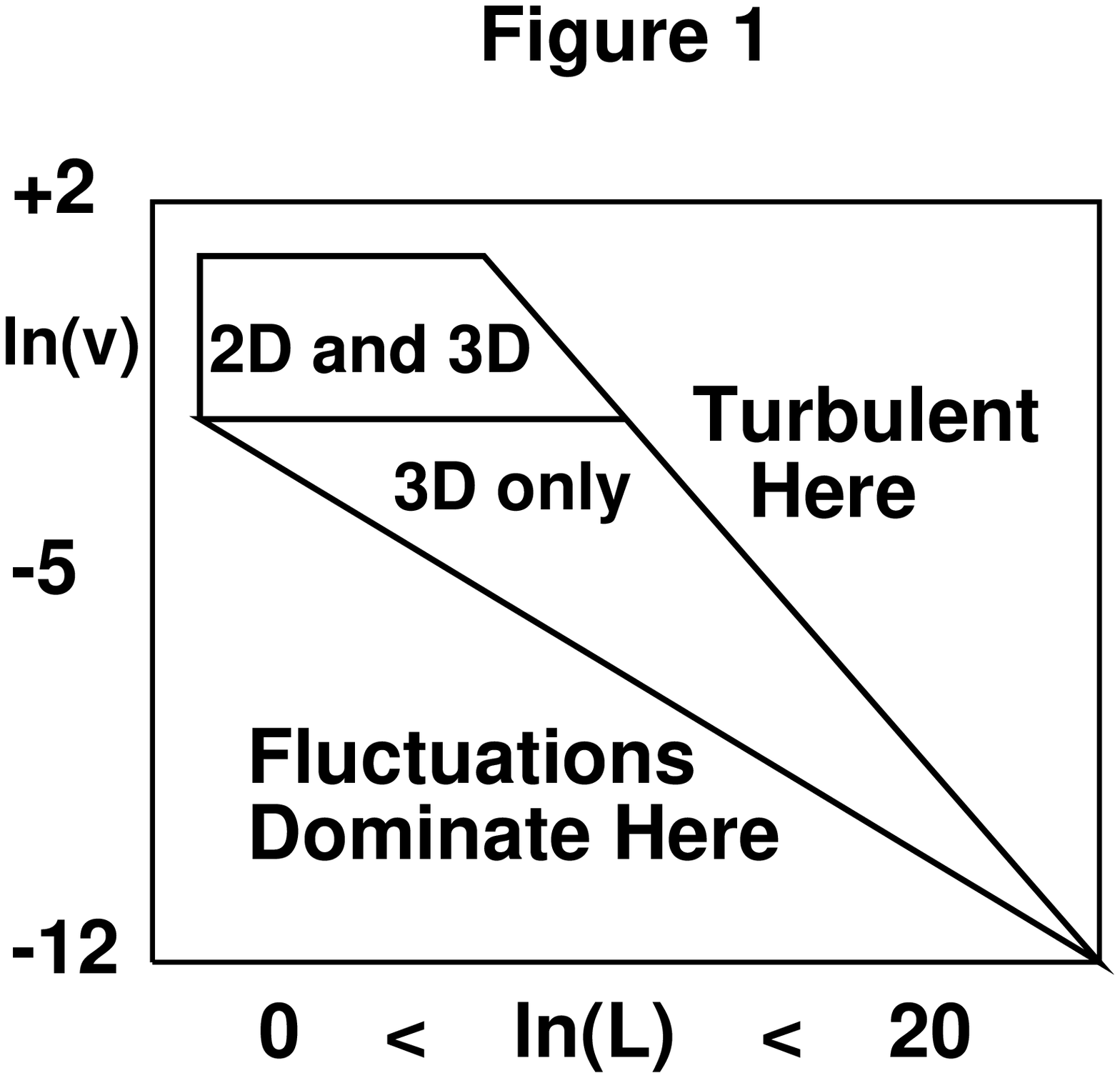}
\caption{
Limits imposed on system size $L$ and boundary velocity $v$ by
atomistic size, thermal fluctuations, turbulence, and shockwaves.  Viscosity
is a useful concept for flows in the enclosed areas.  The figure here is
constructed for a fluid at unit mass, number density, temperature,
viscosity, and heat conductivity.  $L$ is measured in units of the
microscopic particle size and $v$ is measured in units of the thermal
velocity. Modeled after Ref. [6].
}
\end{figure}

Classical fluid flow can be modeled and understood from either the microscopic
or the macroscopic standpoint.  In the {\em microscopic} description individual
particles obey the {\em ordinary} differential equations of motion of
nonequilibrium molecular dynamics\cite{b1}:
$$
\{ \ m \ddot r = m \dot v = F_A + F_B + F_C + F_D  \ \} \ ,
$$
and everything follows from the functional forms of the assumed atomistic,
boundary, constraint, and driving forces\cite{b1,b16}.  In molecular
dynamics, just as in continuum hydrodyamics, the pressure tensor
is (defined to be) the {\em comoving momentum flux}, comoving relative
to the flow velocity.  In a many-body system with forces derived from the
pairwise-additive pair potential $\phi (r)$, the pressure tensor is made
up of $\{ i,j\}$ pair contributions as well as individual particle
convective contributions\cite{b1}:
$$
P_{xy}V = \sum _{i<j}\left( \left[ \frac{xy}{r^2}\right ][F \cdot r]\right )_{ij} + 
\sum _i (p_xp_y/m)_i \ ; \
$$
$$
x_{ij} \equiv x_i - x_j \ ; \ y_{ij} \equiv y_i - y_j \ ; \ r_{ij} \equiv
|r_i - r_j| \ ; \ F_{ij} \equiv -\nabla _i\phi (r_{ij}) \ .
$$
It is evident that the pair-potential pressure tensor is {\em symmetric}, with
$P_{xy}$ and $P_{yx}$ equal.

In the
{\em macroscopic} description the continuum field variables [ such as the mass
density $\rho (r,t)$, the velocity $v(r,t)$, and the energy per unit
mass $e(r,t)$ ] obey {\em partial} differential equations:
$$
\dot \rho = -\rho \nabla \cdot v \ ; 
$$
$$
\rho \ddot r = \rho \dot v = -\nabla \cdot P = \nabla \cdot \sigma \ ; 
$$
$$
\rho \dot e= -\nabla v : P - \nabla \cdot Q \ .
$$
In these general continuum field equations the {\em constitutive relations}
for the stress tensor $\sigma \equiv -P$ and the heat flux vector $Q$
distinguish one material from another.  In both approaches a computational
algorithm for solving the equations is needed.  Its implementation gives
$ \{ \ r(t),v(t) \ \}$ in the microscopic case and
$ \{ \ \rho(r,t),v(r,t),e(r,t) \ \} $ in the  macroscopic case).  In either
case, a well-posed problem also requires boundary and initial conditions,
constraints, and driving forces.

First consider the simplest model system illustrating stationary simple
shear: imagine an incompressible Newtonian fluid, with constant shear
viscosity $\eta $, and which also follows Fourier's linear heat transport
law with a constant heat conductivity $\kappa $:
$$
P_{xy} = -\eta [\ (dv_x/dy) + (dv_y/dx)\ ] \ ;
$$
$$
\rho \dot e = \kappa  \nabla ^2T - P:\nabla v \ .
$$
We ignore thermal expansion, so that the mass density $\rho$ is constant.
We denote the local thermodynamic variables in the conventional way:
temperature $T(r)$, pressure tensor $P(r)$, and internal energy per unit
mass $e(r)$.  We adopt the colon convention in the tensor
product $A:B$ to indicate a sum over all four $A_{ij}B_{ij}$ terms in two
dimensions, and all nine such terms in three dimensions.  In some texts
the alternative sum (immaterial for the symmetric tensors considered here)
$A_{ij}B_{ji}$ is used.

Simple shear flow for this bare bones textbook model is perhaps the simplest
imaginable nonequilibrium flow problem.  It gives a {\em linear} variation
of velocity in space along with a {\em quadratic} variation of temperature.
Simple shear flow can be driven by two moving parallel boundaries, both
of them at temperature $T_B$, and able to absorb heat and to impose
their boundary velocities, $v_x = \pm v$, on a two-dimensional strip or
three-dimensional slab of model fluid of thickness $L$:
$$
v_x(y=\pm L/2) = \dot \epsilon y \ .
$$
The stationary macroscopic description of such a flow, for the model Newtonian
fluid with Fourier heat conduction, has a constant stress tensor, a
{\em linear} velocity profile and (because $\nabla ^2T$ is constant)
a {\em quadratic} temperature profile:
$$
(dv_x/dy) = \dot \epsilon = 2v/L = -P_{xy}/\eta \ ;
$$
$$
\Delta T(y) = T(y) - T_B =
(\eta \dot \epsilon ^2 /2\kappa)[(L/2)-y][(L/2)+y] \ .
$$
The maximum temperature difference, relative to the boundaries' temperature
$T_B$,
$$
\Delta T(y) \leq \Delta T_{\rm max} = \Delta T(0)
= \eta \dot \epsilon ^2L^2/8\kappa = \eta v^2/2\kappa \ ,
$$
occurs at the midplane $y = 0$.  See Fig. 2 for a schematic illustration
of this prototypical simple shear flow.

This stationary solution satisfies {\em energy balance}, with the rate at
which heat is generated throughout the volume $V$ (necessarily the same as
the rate at which external work is done, $-\dot W$) equal to the rate
at which heat is transferred through the two boundary walls of area
(length in two dimensions) $A$, at $y=\pm L/2$:
$$
-\dot W = -P_{xy}V\dot \epsilon = 
\eta V \dot \epsilon ^2 = \kappa A[(dT/dy)_{-L/2}-(dT/dy)_{+L/2}] \ .
$$
This macroscopic description of shear flow guides our interpretation of
microscopic simulations.  We focus on the complications caused by fluctuations
and nonlinearities in what follows.

\begin{figure}
\includegraphics[bb=6.0in 0.0in 8.0in 5.5in,hiresbb=true,angle=-90]{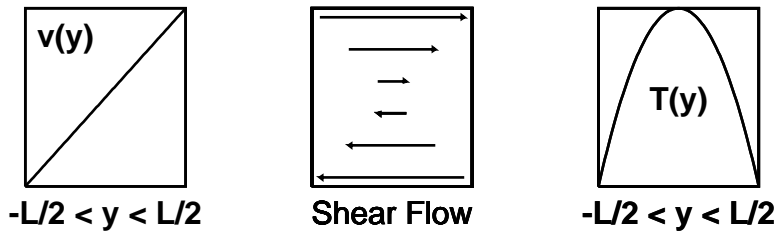}
\caption{
The simple shear flow indicated in the center panel corresponds to the linear
velocity profile (shown to the left) and quadratic temperature profile (shown
at the right).  The nonlinear effects considered in this paper give relatively
small deviations from these idealized profiles.
}
\end{figure}

\section{Microscopic Description of Simple Shear Flow}

Molecular dynamics was developed over 50 years ago \cite{b17,b18,b19}
and soon gave rise to successful interpretations of equilibrium properties
based on hard-sphere perturbation-theory\cite{b20} analogous to Enskog's
hard-sphere understanding of dense-fluid transport properties\cite{b21}.
Field-driven diffusion\cite{b22}, shear and bulk viscous flows\cite{b23},
and heat conducting flows\cite{b24,b25} all came to be simulated with a
variety of algorithms.  Special boundary conditions and computational
thermostats were developed to model these flows.  Throughout this
development the limiting case of Green and Kubo's linear theory of
transport served as a guide\cite{b26}.

In the past thirty years simulation has come a long way.  {\em Billions}
of atoms can be simulated now\cite{b27}.  Nonequilibrium simple-shear
algorithms have been developed for diatomic and polyatomic
molecules\cite{b28,b29},
not just simple fluids.  More recently, alternative shear-flow algorithms
have been developed for periodic irrotational ``elongational''
flows\cite{b30,b31,b32}, flows with a steady stretching in one direction and a
simultaneous shrinking in a perpendicular direction.  It might be
thought (as it once was\cite{b33}) that such a flow could not be followed
forever, but a clever choice of periodic boundaries makes it possible to
study steady thermostated elongational flows.  The conceptual difficulties
involved in deriving these algorithms have led to a spirited
literature\cite{b34,b35,b36} as to the ``correctness'' of the various
algorithms.  Such discussions can easily lead outsiders to the impression
that it is hard to distinguish ``correct'' from ``incorrect'' algorithms.

The controversial aspects of these shear-flow
algorithms\cite{b28,b29,b30,b31,b32,b33,b34,b35,b36} led us to
reconsider the problem.  It seemed to us that a fresh look at the basic
algorithms for monatomic simple-shear flows would help to develop a
perspective clarifying this situation.  It is evident that the nonlinear
aspects of the computer algorithms are to some extent arbitrary, as the
only true guidelines for correctness are consistency with the well-known
and well-accepted {\em linear} flow theory described by Newtonian viscosity
and Fourier heat conduction.

In {\em viscous} shear flow, which we consider here, {\em any}
reasonable algorithm needs to satisfy the requirement that the shear
viscosity for small strain rates agrees with Green and Kubo's linear-response
relation\cite{b2,b26} linking the viscosity to equilibrium fluctuations
in the shear stress
$\sigma_{xy} = - P_{xy}$:
$$
\eta = (V/kT)\int_0^\infty \langle P_{xy}(0)P_{xy}(t)\rangle _{\rm eq}dt \ ;
$$
$$
\eta = \frac{-P_{xy}}{\dot \epsilon} \ ; \ \dot \epsilon \equiv \
\frac{\partial v_x}{\partial y} + \frac{\partial v_y}{\partial x} \ .
$$

For simplicity we consider systems with pairwise-additive forces
$\{ F_{ij} \}$ derived from a potential function $\Phi = \sum \phi $:
$$
\{ \ F_{ij} = -\nabla _i \phi (|r_i - r_j|) = -F_{ji} \ \} \ ,
$$
In such a system both the energy $E$,
$$
E = \Phi + K = \sum _{i<j} \phi _{ij} + \sum _i p_i^2/2m \ ,
$$
and the microscopic pressure tensor $P$,
$$
PV = \sum _{i<j}(Fr)_{ij} + \sum _i(pp/m)_i \ ,
$$
are sums of two-particle potential and single-particle kinetic
contributions.

There are two interesting ambiguities in the definition of pressure
in a particulate system.  The contributions of the potential pair
interaction terms $\{ \ (Fr)_{ij} \ \}$ to $PV$ need to be allocated spatially.
Either delta-function contributions at $r_i$ and $r_j$, or at
$(r_i+r_j)/2$, or {\em smoothed} distributions centered at these
locations can be used. The old ``Irving-Kirkwood'' preference for
delta functions is motivated more by analytic convenience than by
any physical considerations. A {\em smoothed} approach is certainly
preferable in computational work.

The kinetic part of $PV$ is
ambiguous too.  How is the ``comoving'' velocity to be defined in
a system with transient velocity fluctuations?  Either an
instantaneous or a time-averaged hydrodynamic velocity can be chosen.
Only the {\em kinetic} part of $P_{xy}V$ is affected by this choice.  The
simplest versions of the two choices are as follows:
$$
P_{xy}^K(t)V \equiv \sum _i
m[v_x^i -  v_x(t)]
 [v_y^i -  v_y(t)] \ ({\rm instantaneous})
$$
or
$$
P_{xy}^K(t)V \equiv \sum _i
m[v_x^i - \langle v_x \rangle _{\rm time}]
 [v_y^i - \langle v_y \rangle _{\rm time}] \ ({\rm time}\ {\rm averaged}) \ .
$$
The kinetic part of the pressure tensor is the usual definition of the
instantaneous kinetic {\em temperature} for $N$ particles\cite{b37}.  In
three space dimensions the usual definitions of
$\{ T_{xx},T_{yy},T_{zz} \} $ are:
$$
NkT_{xx}(t) = \sum ^N m(v_x^i-\langle v_x(t)\rangle)^2 \ ; \
$$
$$
NkT_{yy}(t) = \sum ^N m(v_y^i-\langle v_y(t)\rangle)^2 \ ; \
$$
$$
NkT_{zz}(t) = \sum ^N m(v_z^i-\langle v_z(t)\rangle)^2 \ ,
$$
where the average velocity at time $t$ is computed by summing $N$ individual
particle velocities.

The instantaneous hydrodynamic velocity $(v_x(t),v_y(t))$ has to
be estimated numerically.  The simplest way to do this in a system
composed of moving particles, is to use a weighted (smoothed-particle)
average of nearby particle velocities:
$$
v(t) \equiv \sum _i v_i w_{ir} / \sum _i w_{ir} \ ; \
w_{ir} = w(|r - r_i|) \ ,
$$
where the time-independent scalar weight function $w_{ir}$ has a
finite spatial range, and emphasizes the contributions of those
nearby particles to the location $r$ where $v(t)$ is to be computed.  The
details of this spatial average are described in Section VII.  Fortunately,
the main conclusions of this work are independent of the necessarily
arbitrary choice between using instantaneous and time-averaged flow
velocities.  By choosing to study stationary simple shear we avoid the
additional complexities associated with {\em time} averaging flows which
have an explicit time dependence in their boundary conditions.

The atomistic simulation approach is also complicated by heat and by
fluctuations.  Because viscosity is {\em dissipative}, a viscous shear
flow necessarily generates {\em heat}.  Particle motions and collisions
in systems tractable with molecular dynamics necessarily exhibit
fluctuations in {\em all} their local properties\cite{b38}.  Despite the
time-reversibility of the underlying equations of motion, the overall
longtime-averaged character of these flows {\em is} necessarily
dissipative, in accord with macroscopic hydrodynamics and the Second
Law of Thermodynamics\cite{b38,b39,b40,b41}.

To simulate a {\em steady} viscous flow, the heat generated needs to
be removed.  This can be done with any one of many schemes, all of which
are based on time-reversible
constraint forces\cite{b1,b16}.  Such constraint forces can play the r\^ole of
a feedback-based thermostat or ergostat,
$$
\{ \ F_{\rm Constraint} = -\zeta p \ \} \rightarrow
\dot K \equiv 0 \ {\rm or} \ \dot E \equiv 0 \ .
$$
Typical constraint-force choices keep the kinetic energy $K$ or the
total energy $E$ fixed, or allow these energies to fluctuate about a
specified mean value in a way consistent with Gibbs' statistical
mechanics at equilibrium\cite{b42,b43}.  For simplicity, we restrict
ourselves here to ``Gaussian thermostats'' [ so named after Gauss' Principle
of Least Constraint ]\cite{b44} which fix the kinetic energy $K$ or the
total energy $E$ of a particular set of degrees of freedom by imposing
feedback-based constraint forces, $\{ \ F_{\rm Constraint}  = -\zeta p \ \}$.
The Gaussian ``friction coefficient'' $\zeta $ constrains the momenta
$\{ \ p \ \}$ contributing to $K$ or $E$.  We next describe the two
best-known computational algorithms for simulating simple shear.

\section{Doll's and Sllod Algorithms for Simple Shear Flow}

Shear flows were the first application of {\em homogeneous} nonequilibrium
molecular dynamics\cite{b6}.  Steady {\em inhomogeneous}
shockwaves\cite{b45} and
boundary-driven shear flows\cite{b6,b7,b8,b9} had both been simulated earlier.
Special boundary conditions, or external forces, as well as thermostats
had to be developed for the homogeneous shear flows.  The numerical
applications of these ideas, in the early 1970s, preceded their formal
theoretical development by decades.  The Doll's Tensor Hamiltonian for
viscous flows was discovered in 1985\cite{b4}; Dettmann and Morriss'
Hamiltonians (for the Nos\'e-Hoover and Gauss thermostats) were
discovered in 1996\cite{b46,b47}; and a proper formulation of elongational
flows first appeared in 1998\cite{b30}.

Simple homogeneous shear flow, with the $x$ velocity proportional to
the $y$ coordinate,
$$
\dot \epsilon \equiv dv_x/dy \ \longleftrightarrow \
\nabla v \equiv \
\left[
\begin{array}{cc}
\frac{\partial v_x}{\partial x}   &  \frac{\partial v_y}{\partial x}  \\
\frac{\partial v_x}{\partial y}   &  \frac{\partial v_y}{\partial y}       
\end{array}
\right] \ = \
\left[
\begin{array}{cc}
0               &  0  \\
\dot \epsilon   &  0       
\end{array}
\right] \
,
$$
can be implemented with periodic ``Lees-Edwards'' boundary
conditions\cite{b48} [ developed independently by Bill Ashurst in his
Ph D thesis work ]\cite{b7}.  See page 26 of Ref. [6] for a brief
description of Ashurst's algorithm.  The corresponding flow is illustrated
in Figure 3.

\begin{figure}
\includegraphics[height=9cm,width=8cm,angle=-90]{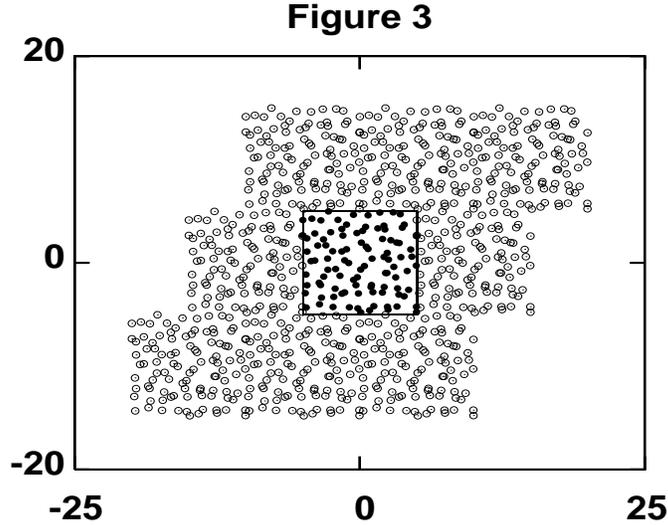}
\caption{
Periodic 100-particle shear flow, showing eight periodic
$L \times L$ images of the central
$10 \times 10$ box.  The particle velocities in the images above/below the
central $N$-particle box differ from those of the central box by
$\pm \dot \epsilon L$.
}
\end{figure}

This simple shear flow, with $\frac{\partial v_x}{\partial y}$
nonzero, is ``rotational'', in the sense that the clockwise rotation rate,
$-\omega$, is nonzero, and equal to half the strain rate:
$$
-\omega \equiv [(\partial v_x/\partial y) - (\partial v_y/\partial x)]/2 =
\dot \epsilon /2 \ .
$$
In our specific special case $\nabla v$ can be written as a sum of irrotational
and rotational contributions:
$$
\nabla v = 
\left[
\begin{array}{cc}
0               &  0  \\
\dot \epsilon   &  0       
\end{array}
\right] \ =
\left[
\begin{array}{cc}
0               &  +\dot \epsilon /2  \\
+\dot \epsilon/2   &  0       
\end{array}
\right] \ +
\left[
\begin{array}{cc}
0               &  -\dot \epsilon /2  \\
+\dot \epsilon /2  &  0       
\end{array}
\right] \ .
$$
See Figure 3 for an illustration of the periodic boundary conditions consistent
with such a flow.  A stationary nonequilibrium state can result if
the irreversible heating rate, $-\dot W = \eta \dot \epsilon ^2 V$ is
appropriately compensated by momentum-dependent thermostat forces.

The Doll's and Sllod algorithms are straightforward possibilities for
simulating such a flow, and the differences between them are small
(of order $\dot \epsilon ^2$) at moderate strain rates.  Ashurst's
periodic shear algorithm preceded the formal derivations of differential
evolution equations for the shear flow and its associated thermostats.  His
finite-difference algorithm was equivalent, for small timesteps, to a
thermostated version of what is now called the Sllod algorithm:
$$
\{ \
\dot x = (p_x/m) + \dot \epsilon y \ ; \ \dot y = (p_y/m) \ ; \
\dot p_x = F_x - \dot \epsilon p_y - \zeta p_x \ ;
\ \dot p_y = F_y - \zeta p_y \ \} \ .
$$
In the absence of the thermostat forces $\{ \ -\zeta p \ \}$, these
motion equations give an energy change exactly consistent with the
rate at which thermodynamic work is performed by the instantaneous
shear stress $-P_{xy}$.

In the alternative Doll's-Tensor approach\cite{b4}, a coordinate-momentum
$(q,p)$ sum [ giving rise to the ``Kewpie-Doll'' name ] is added to the
usual Hamiltonian.  The added term includes the strain rate
$\dot \epsilon $:
$$
 {\cal H}_{\rm Doll } = {\cal H}_{\rm Usual } +
\dot\epsilon  \sum _i (yp_x)_i \ .
$$
The resulting Hamiltonian equations of motion,
$$
\{ \ \dot q \equiv +\frac{\partial {\cal H}}{\partial p} \ ; \
\dot p \equiv -\frac{\partial {\cal H}}{\partial q} \ \} \ ,
$$
are only slightly different to the closely related ``Sllod'' equations.
The Doll's tensor motion equations are
$$
\{ \
\dot x = (p_x/m) + \dot \epsilon y \ ; \ \dot y = (p_y/m) \ ; \
\dot p_x = F_x \ ; \ \dot p_y = F_y - \dot \epsilon p_x \ \} \ .
$$
In the absence of heat-extracting thermostats, both approaches, Sllod
and Doll's, provide an energy change exactly consistent
with thermodynamics:
$$
\dot {\cal H}_{\rm Doll }(q,p,\nabla v) = 0 \rightarrow
\dot {\cal H}_{\rm Usual}(q,p) = -(d/dt) \sum qp:\nabla v \equiv
-\dot \epsilon P_{xy}V \ .
$$

The extra momentum-dependent forces,  $\{ -\dot \epsilon p_y\} $ for
Sllod and  $\{ -\dot \epsilon p_x\} $ for Doll's, model rotation.
In a rotational flow in the $xy$ plane the rotation rate $\omega $
is given by
$$
\omega \equiv \frac{1}{2}\left[\frac{\partial v_y}{\partial x}
            - \frac{\partial v_x}{\partial y}\right] \ ,
$$
and {\em corotating} momenta would include rotational contributions:
$$
\dot p_x = -\omega p_y \ ; \ \dot p_y = +\omega p_x \ .
$$
The Coriolis accelerations corresponding to rotation are similar
in form to the momentum-dependent forces included in the Doll's and
Sllod algorithms.

It should be noted that the ambiguity in treating the atomistic
rotation associated with shear has a relatively well-known analog in
continuum mechanics\cite{b49}.  In the plastic flow of solids the total
macroscopic strain, idealized as a sum of elastic and plastic contributions, can
only be computed by time integration:
$\int \dot \epsilon dt \rightarrow \epsilon$.  Because stress depends on
strain, calculations involving plastic flow require the simultaneous
time integration of
an equation for stress evolution, $\int \dot \sigma dt \rightarrow \sigma $.
If the effect of rotation on the stress is included (imagining that the stress
rotates as if it were embedded in a rigid body) the ``Jaumann stresses''
result.  These laboratory-frame stresses satisfy the evolution equations:
$$
\dot \sigma _{xx} = -2\omega \sigma _{xy} \ ; \
\dot \sigma _{yy} = +2\omega \sigma _{xy} \ ; \
\dot \sigma _{xy} =  \omega [\sigma _{xx} - \sigma _{yy}] \ .
$$

A more complicated boundary-driven shear flow can be driven by imposing
appropriate boundary conditions, including external thermostats or ergostats,
 on a Newtonian region with the usual equations of motion:
$$
\{ \ \dot x = (p_x/m) \ ; \ \dot y = (p_y/m) \ ; \
\dot p_x = F_x \ ; \ \dot p_y = F_y \ \} \ .
$$
The details of our boundary-driven simulations are given in Sec. VI.  In
his thesis work in the early 1970s
Ashurst\cite{b7} found consistent shear viscosities for both
boundary-driven and homogeneous periodic shear flows.  More recently
Liem, Brown, and Clarke  made an effort to characterize the very
nonlinear effects we seek in the present work in shear flow.  They compared
homogeneous shear (presumably using the Sllod approach) with a very large scale
boundary-driven flow\cite{b9}.  Unfortunately the limitations inherent in
using such large systems made the nonlinear effects too small to measure.

\section{Boundary-Driven Shear Flow}

\begin{figure}
\includegraphics[height=12cm,width=10cm,angle=0]{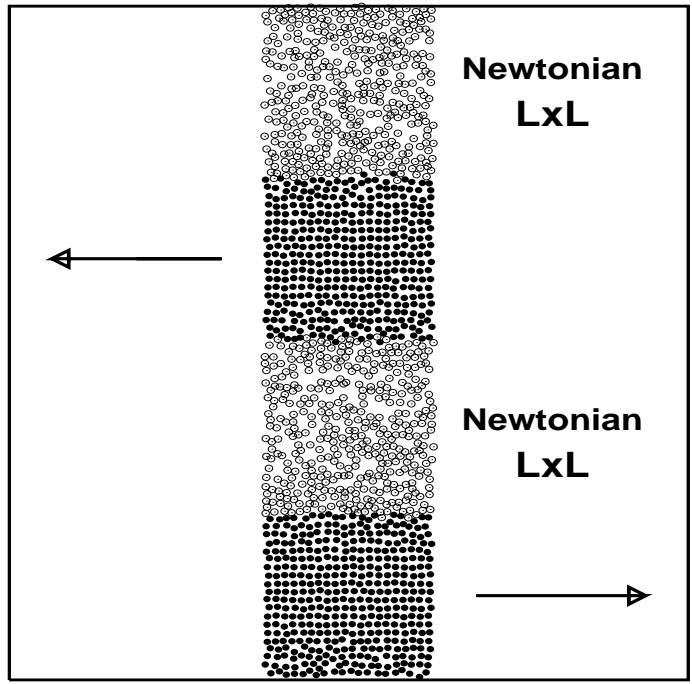}
\caption{
Four-chamber $L \times 4L$ periodic flow with $4 \times 400$ particles.
Chamber 1 (filled circles at the bottom)
moves to the right at speed $+\dot \epsilon  L/2$; chamber 3 (filled
circles, third chamber from the bottom) moves to the left at speed
$-\dot \epsilon  L/2$.  This imposes the nominal strain rates
$\mp \dot \epsilon$ on the two Newtonian chambers 2 and 4 (open circles).
Periodic boundaries
apply in both the $x$ and the $y$ directions.  Note the ordering effect
of the moving tethers in Chambers 1 and 3.
}
\end{figure}

To avoid dealing with free surfaces we have chosen to simulate
fully-periodic four-chamber systems of the type shown in Figure 4.
Posch and Hoover used the same geometry in an investigation of heat
flow\cite{b50}.  In two dimensions the system
is made up of four $L \times L$ cells, each with an area one-fourth the
total, $L^2 = N/4$.  To eliminate surface effects the system is periodic
in both $x$ and $y$, with two steadily-moving chambers thermostated at a
fixed kinetic temperature.  The motion is induced by using steadily moving
tethers.  The remaining particles obey Newton's
motion equations.  This four-chamber geometry has a statistical advantage.
The two
Newtonian regions provide independent estimates of the stress and
temperature tensor profiles.  The differences between the two Newtonian
regions provides a useful indication of the estimates' reliability.  See,
for instance, the normal-stress profiles for the two regions shown in
Figs. 8, 10, and 11.

Though we do not give the results here, we have also considered a
{\em two}-chamber periodically shearing flow geometry, in both two and
three space dimensions.  In the two-chamber approach the moving tethers'
$x$ velocities are
$$
\{ \
\pm 1\dot\epsilon L \ , \
\pm 3\dot\epsilon L \ ; \
\pm 5\dot\epsilon L \ ; \ \dots \} \ ,
$$
in the odd-numbered cells, where $L$ is the cell width, while the motion
in the even-numbered cells is purely Newtonian.  Just as in the
four-chamber case periodic shearing boundaries are used in all the
space dimensions.  The results from the two-chamber and four-chamber
flows are not significantly different and are therefore not elaborated
here.

The multi-chamber shear flows are
induced by {\em tethering} the boundary particles to steadily moving
lattice sites (a moving square lattice or simple cubic lattice, for
convenience) with a simple quartic tethering
potential, just as in the $\phi^4$ model for heat conduction\cite{b51,b52}:
$$
\phi^4 = \frac{\kappa _4}{4}[r-r_0(t)]^4 \ ; \
\dot r_0(t) = (\pm \dot \epsilon L/2,0) \ .
$$
Here the sites move.
100 is a good choice for the force constant $\kappa _4$ and gives
better equilibration than tethers with alternative power-law exponents.

In order to prevent diffusive mixing between the Newtonian
particles and the boundary particles it is useful to include a smooth
repulsive boundary force returning errant Newtonian particles toward their
chambers.  For convenience this force has the same form as the boundary
tethering forces used in the moving chambers.
The fixed speed of the tethers, $|\dot r_0(t)|$, is chosen to impose
overall strain rates $\pm \dot \epsilon$ on the two Newtonian regions.
Whether or not the strain rate actually penetrates into the Newtonian regions
depends upon whether or not ``slip'' occurs at the boundary, as we 
discuss in Sec. VIIIB.

Because the heat generated by the shear varies as $L^D$ in $D$ dimensions,
but must flow across a boundary of area (or length) $L^{D-1}$ something
``interesting'' necessarily occurs as $L$ increases.  Eventually the
driving chambers must become decoupled from the bulk Newtonian chambers
so that both the work done and the corresponding heat generated actually
decrease with increasing strain rate. In favorable cases, with
good frictional coupling at the boundaries, this driving technique provides good
velocity profiles and allows nonlinear effects to be determined.  We
turn next to the instantaneous smooth-particle spatial-averaging procedure
required for
analyzing the simulations.

\section{Smooth-Particle Spatial Averages}

The spatial averaging algorithms our simulations require are borrowed from
a continuum technique, ``SPAM''\cite{b3}.  Smooth particle applied
mechanics (``SPAM'') is a technique for solving the partial differential
continuum equations (continuity, motion, and energy) for the evolution
of the density, velocity, and energy.  It makes use of a normalized weight
function with a maximum range $h$, $w(|r|<h)$.  The weight function describes
the spatial extent of a representative particle of mass $m$.  The weight
function is formulated with at least two continuous derivatives everywhere in order
that the first and second continuum derivatives of smooth particle sums
[ corresponding to instantaneous local quantities such as
$\nabla \rho (r,t)$, $\nabla v(r,t)$, and $\nabla ^2T(r,t)$ ], are
everywhere continuous in both space and time.  This continuity and
differentiability facilitates a smooth transition between particle and
continuum analyses.

Consider the two simplest cases, mass and momentum
sums.  The density $\rho(r)$, and the momentum density $\rho (r) v(r)$ at the
location $r$ are {\em defined} by summing up the contributions of all
particles $\{j\}$ within the maximum range $h$ of that location:
$$
\rho(r) \equiv m \sum_jw(|r-r_j|) = m \sum w_{rj} \ ; \
$$
$$
\rho(r)v(r) \equiv m \sum_jv_jw(|r-r_j|) = m \sum w_{rj}v_j \ .
$$
An advantage of this formulation is that the continuum continuity equation,
$$
\frac{\partial \rho}{\partial t} = -\nabla \cdot (\rho v) \ ,
$$
is satisfied {\em exactly} by the interpolated smooth particle fields:
$$
\left(\frac{\partial \rho(r)}{\partial t}\right)_r =
m \sum_j\left(\frac{\partial w(|r-r_j|)}{\partial t}\right)_r =
$$
$$
+m \sum_j(\nabla _j w_{rj})\cdot
\left(\frac{\partial |r-r_j|}{\partial t}\right)_r =
+m \sum_j(\nabla _j w_{rj})\cdot v_j =
$$
$$
-m\nabla _r \sum w_{rj}v_j \equiv
-\nabla _r\cdot (\rho v) \ .
$$
Notice that the spatial gradient operator, $\nabla _r$, affects only
$w(r-r_j)$ and not the individual point-particle properties $\{ v_j\}$.

By choosing properly normalized weight functions spatial averages can
be computed in one, two, or three spatial dimensions.  The pressure tensor
in the vicinity of a particle or at a grid point in three dimensions would
be computed by using the {\em three}-dimensional weight function:
$$
w_{3D}(r) = (105/16\pi h^3)[1 - 6(r/h)^2 + 8(r/h)^3 - 3(r/h)^4] \rightarrow
\int_0^h 4\pi r^2w_{3D}(r)dr \equiv 1 \ .
$$
This particular weight function was discovered and used by Lucy in
1977\cite{b53}.
It is the simplest normalized polynomial with (1) a maximum at $r=0$; (2)
a finite range $r<h$, and (3) two continuous derivatives vanishing at $r=h$.
The corresponding one- and two-dimensional weight functions (for averages
in thin strips or slabs, and for averages at particles or grid points in
two-dimensional problems, are
$$
w_{2D}(r) = (5/\pi h^2)[1 - 6(r/h)^2 + 8(r/h)^3 - 3(r/h)^4] \rightarrow
\int_0^h 2\pi r w_{2D}(r)dr \equiv 1 \ .
$$
$$
w_{1D}(r) = (5/4 h)[1 - 6(r/h)^2 + 8(r/h)^3 - 3(r/h)^4] \rightarrow
\int_0^h 2 w_{1D}(r)dr \equiv 1 \ .
$$
The smooth-particle equations of motion for the time development of the
individual particle velocities $\{v_j\}$ are generally formulated so as
to conserve linear momentum exactly.  The failure of this approach to
conserve {\em angular} momentum is a relatively subtle point worthy of
more research investigation\cite{b54}.

\section{Numerical Results: Two Dimensions}

\subsection{Homogeneous Simulation Results with Lucy's Potential}

To simplify the numerical work, reduce numerical integration errors, and
to make contact with existing data\cite{b55,b56,b57}, we initially chose
to use Lucy's pair
potential in two space dimensions.  This pairwise-additive potential is
identical to the weight function $w_{2D}(r)$ of the previous Sec. VI:
$$
\phi (r) = \frac{5}{\pi h^2}[1 - 6x^2 + 8x^3 - 3x^4] \ ; \
x \equiv \frac{r}{h} < 1 .
$$

We can make a rough estimate of the potential contribution to the
energy and pressure for this potential by assuming a {\em random} distribution
of particles.  Viewed as a statistical-mechanical potential function for
molecular dynamics in two dimensions, Lucy's potential (due to the
normalization of the Lucy weight function) then corresponds
to the simple equation of state for a two-dimensional ideal gas:
$$
\Phi = \sum _{i<j} \phi _{ij} = \frac{N^2}{2V} \ \leftrightarrow
\ P^\Phi V = \frac{N^2}{2V} \ .
$$
The energy expression follows from the normalization condition, while
the pressure expression follows from the virial theorem:
$$
P^\Phi V = \frac{1}{2}\sum r_{ij} \cdot F_{ij} \ \simeq \
-\frac{N^2}{4V}\int _0^h2\pi r^2\phi^\prime (r)dr \ = \ \frac{N^2}{2V} \ .
$$
At unit mass and number density and at
a strain rate of $\dot \epsilon = 0.05$ the shear viscosity and normal
stresses (for Sllod) were measured precisely in 1995\cite{b55,b56}.  Because
the results are insensitive to the number of particles used we list in
Table I below only a single representative set of simulations for $N=400$.
The insensitivity of the viscosity to the algorithm type is also evident
in Daivis recent work\cite{b58}.

We chose $h=3$, for which the Sllod algorithm shear viscosity (with a
single thermostat variable) has previously been computed.
At unit density, with $L \times L = N$, the average number of pair
interactions is approximately $14N$.  We also used exactly the {\em same}
Lucy's function as a {\em weight function}\cite{b3} for carrying out
spatial averages. Table I lists the pressure-tensor components using
both the Doll's and Sllod algorithms.  The first four simulations are
``isothermal'' with both the average temperature, $(T_{xx}+T_{yy})/2$
thermostated (S1 and D1), and with both temperatures separately
thermostated (S2 and  D2), by using two control variables, $\zeta _x$
and $\zeta _y$.  The last two simulations (SE and DE) constrain the
internal energy rather than the temperature.  Notice that in all these
simulations the average pressure is fairly close to the simple
virial-theorem estimate, $P^\Phi V/N \simeq 0.5$.

The normal stress difference is sensitive to the single-thermostat type:
$$
P_{xx} - P_{yy} = + 0.015 \ ({\rm Sllod}) \ ; \
P_{xx} - P_{yy} = - 0.017 \ ({\rm Doll's}) \ ,
$$
and is almost entirely kinetic.  At the high effective density of these
simulations the potential portion of the stress distribution is nearly
isotropic.  These results are essentially unchanged
if energy, rather than average temperature, is controlled.  On the other
hand, this striking normal-stress effect disappears completely (because
it is kinetic in nature) if the two temperatures are {\em separately}
thermostated.  We conclude from these simulations that a definitive
determination of the normal
stress difference requires a more realistic boundary-driven flow.  One
can have no confidence in nonlinear kinetic effects when the evolving
kinetic energy itself can be dominated by the choice of homogeneous
thermostats, one or two.

\hspace{0.5 cm}

{\bf Table I}.  Relatively {\em high-density} Lucy viscosity in two
dimensions.  The mass and number densities are unity:
$\rho = mN/V = N/V = 1$.  The {\em range} of the Lucy potential is
$h=3$, so that each particle interacts simultaneously with approximately 30 neighbors. 
Space-and-time-averaged pressure tensors are given here for homogeneous Sllod and
Doll's algorithms with a square periodic $20 \times 20 = 400$-particle
cell with a strain rate of $\dot \epsilon = dv_x/dy = 0.05$.
The kinetic temperature is fixed, $kT = 0.07$, using a single control
variable $\zeta$ (in the two runs S1 and D1) and two separate control
variables in the $x$ and $y$ directions, $\zeta _x$ and $\zeta _y$
(in the two runs S2 and D2).  The total energy is fixed at 0.500 for
the runs SE and DE.  The pressure-tensor components are given in the
order $xx,yy,xy$ with the kinetic, potential, and total terms indicated.
The boundary conditions are periodic, with a total run time of
$200\times 10,000$ timesteps.  The fourth-order Runge-Kutta timestep is
0.005. The average potential energies per particle for the first four
runs are all in the range $0.4272 < \Phi /N < 0.4274$ so that the total
energy in these runs is 0.497.

\hspace{0.5 cm}

\begin{tabular}{| c || ccc || ccc || ccc |}
        \hline
Run Type  & $P_{xx}^K$ &$ P_{xx}^\Phi$ & $P_{xx}^\Sigma$ &
                  $P_{yy}^K$ &$ P_{yy}^\Phi$ & $P_{yy}^\Sigma$ &
                  $P_{xy}^K$ &$ P_{xy}^\Phi$ & $P_{xy}^\Sigma$  \\
\hline
 400S1&0.078&0.484&0.562&0.062&0.485&0.547&-0.022&+0.002&-0.021 \\
 400D1&0.060&0.486&0.546&0.079&0.483&0.563&-0.021&+0.001&-0.020 \\
 400S2&0.070&0.485&0.555&0.070&0.484&0.554&-0.025&+0.002&-0.023 \\
 400D2&0.070&0.485&0.555&0.070&0.484&0.554&-0.025&+0.002&-0.023 \\
 400SE&0.081&0.484&0.564&0.064&0.485&0.549&-0.023&+0.002&-0.022 \\
 400DE&0.062&0.486&0.548&0.083&0.483&0.566&-0.022&+0.001&-0.021 \\
\hline
\end{tabular}

\subsection{Boundary-Driven Simulation Results with Lucy's Potential}

\begin{figure}
\includegraphics[height=6cm,width=6cm,angle=0]{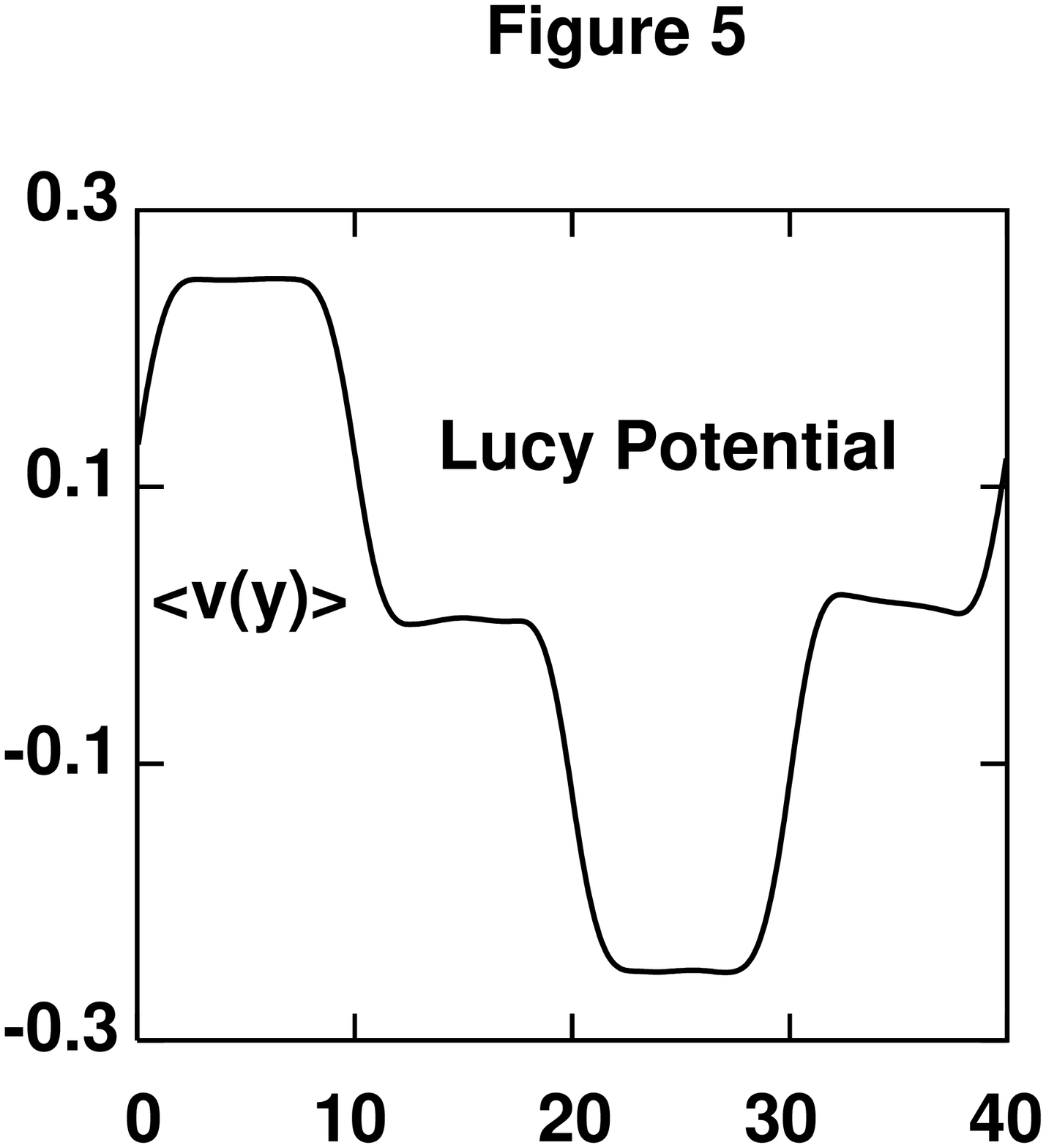}
\caption{
400-particle Lucy velocity profile.  See Fig. 4 for a 1600-particle
analog.  The two moving chambers correspond to the maximum and the
minimum velocities.  Here the two Newtonian chambers are so poorly
coupled to the moving chambers that their mean velocities are nearly
zero.
}
\end{figure}

Because the homogeneous investigations with the Lucy potential were
sensitive to thermostat type, we next carried out several 
four-chamber simulations with two boundary chambers moving oppositely,
as shown in Fig. 4.  The density and strain rate were chosen
to match the data in Table I.  These simulations produced no useful 
normal-stress results!
This failure reflects the nearly negligible coupling between the
two moving chambers and the two Newtonian chambers. See the typical
velocity profile in Fig. 5.  Particles in the two Newtonian chambers
simply rest quietly between the two rapidly moving walls.
Evidently this very dense repulsive fluid with relatively
weak collisonal forces has insufficient friction for boundary driving
to reach strain rates with significant nonlinear stress differences.
This same difficulty was found by Liem, Brown, and Clarke in their
three-dimensional Lennard-Jones simulations\cite{b9}.

On the other hand, in Ashurst's thesis work his ``fluid-wall'' boundary
driving regions (velocity constraints, but no tethers, were used)
produced good linear velocity profiles for both Lennard-Jones and
soft-sphere potentials\cite{b6,b7}.  Because
those simulations correspond to a much lower density (with about three
interactions per Lucy particle rather than thirty) and much more
violent collisions than those of Table I we abandoned the high-density
Lucy simulations and took up instead soft-disk and soft-sphere simulations at
conditions more closely resembling those of Ashurst.  The forcelaw change
was motivated by the desire to check our results with those from
previous simulations\cite{b57}.  The new simulations are
discussed in the following two subsections.

\subsection{Homogeneous Simulation Results with a Soft Disk
  Potential}

\begin{figure}
\includegraphics[height=6cm,width=6cm,angle=-90]{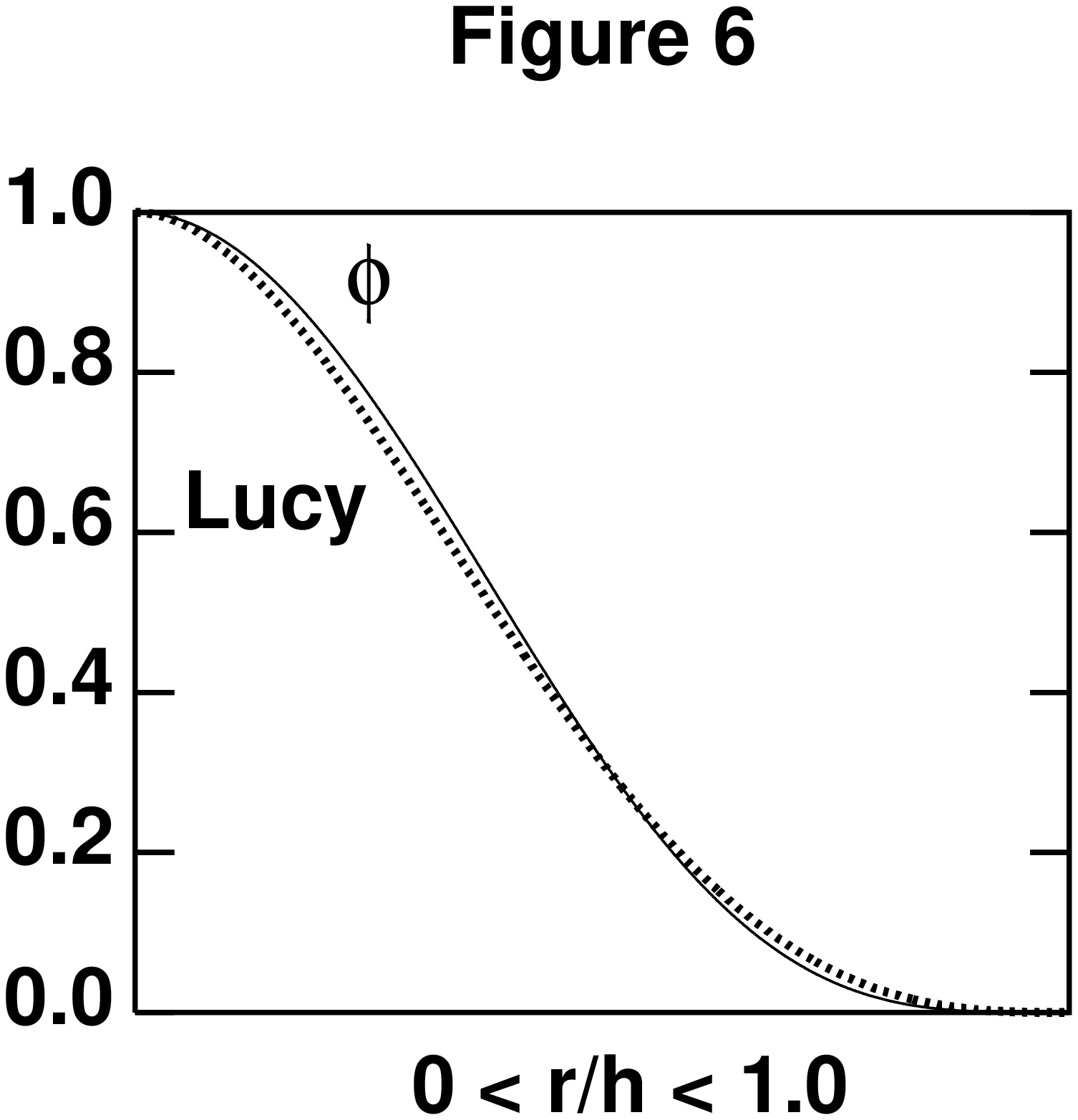}
\caption{
Comparison of Lucy's potential and the short-ranged soft-sphere
repulsive potential $\phi $.  The ordinate and the abscissa are
divided by their maximum values, so that both scales vary from 0 to 1.
}
\end{figure}

Here we consider the short-ranged smooth soft-disk repulsive
potential\cite{b57},
$$
\phi (r<1) = 100(1 - r^2)^4 \ ; \ r^2 = x^2 + y^2 \ ,
$$
very similar in general shape to Lucy's, but with {\em three} vanishing
derivatives rather than just two, at its maximum range of unity.
The two potentials are compared in Fig. 6.  In comparison simulations
we found that the two potentials provide quite similar normal stresses
at corresponding temperatures and densities.  Here we
choose unit density and energy per particle.  We will see that
boundary-driven simulations with this potential choice
provide good velocity profiles, as did Ashurst's similar ``fluid walls'' in 1972,
and also allow comparisons with previous viscosity simulations in
the same thermodynamic state.  First we consider again Sllod and Doll's
simulations with
homogeneous shear.

Just as in the Lucy simulations, the soft-disk normal stresses are quite
different for the Doll's and Sllod algorithms and are even less sensitive to
system size.  Now the short-ranged potential contribution to shear stress
is actually
dominant.  The results listed in Table II, all using a single
friction coefficient, fixing the energy per particle, $E/N = 1$,
show that now the potential
contribution to the stress difference is the same order as the kinetic
contribution.  The Sllod pressure difference is considerably smaller
in magnitude:
$$
P_{xx} - P_{yy} = + 0.008 \ ({\rm Sllod}) \ ; \
P_{xx} - P_{yy} = - 0.054 \ ({\rm Doll's}) \ ,
$$
Because the rotated normal
stress difference is equivalent to a shear stress:
$$
\frac{(P_{xx}-P_{yy})}{2} \stackrel{45^o}{\longleftrightarrow} P_{xy} \ ,
$$
an alternative
phenomenological description of the normal-stress effect corresponds to a rotation
of the principal stress direction (the direction of maximum tension
in the traceless stress tensor).  The Doll's algorithm gives a
clockwise shear-stress rotation while the Sllod rotation is counterclockwise.

\hspace{0.5 cm}

{\bf Table II}.  Moderate-density soft-disk viscosities. The
range of the potential is unity so that each particle interacts, on
the average, with only three neighbors at unit density.
Space-and-time-averaged pressure tensors for homogeneous
Sllod and Doll's algorithms using the soft-disk pair
potential illustrated in Fig. 6, $\phi = 100(1-r^2)^4$.
The energy and density are fixed, and equal to unity, and the strain rate,
$\dot \epsilon = dv_x/dy$ is 0.50.  The pressure-tensor components,
are again given in the order $xx,yy,xy$ with the kinetic, potential, and
total terms indicated.  The boundary conditions are periodic, with a
total run time of 200x10,000 timesteps for N = 64, 256, and 1024;
200x2000 timesteps for N = 4096, and 200x500 timesteps for N = 16,384.
The fourth-order Runge-Kutta timestep is 0.005 and a single isoenergetic
friction coefficient is used, as explained in Sec. II of the text.

\hspace{0.5 cm}

\begin{tabular}{| c || ccc || ccc || ccc |}
        \hline
N Type & $P_{xx}^K$ &$ P_{xx}^\Phi$ & $P_{xx}^\Sigma$ &
       $P_{yy}^K$ &$ P_{yy}^\Phi$ & $P_{yy}^\Sigma$ &
       $P_{xy}^K$ &$ P_{xy}^\Phi$ & $P_{xy}^\Sigma$ \\
\hline
   64S&0.690 &3.217 &3.907& 0.681& 3.223&3.904& -0.081& -0.534& -0.615 \\
   64D&0.679 &3.201 &3.880& 0.692& 3.240&3.932& -0.081& -0.535& -0.617 \\
  256S&0.691 &3.214 &3.904& 0.682& 3.216&3.898& -0.084& -0.538& -0.622 \\ 
  256D&0.679 &3.195 &3.873& 0.694& 3.235&3.929& -0.085& -0.538& -0.623 \\ 
 1024S&0.691 &3.212 &3.904& 0.681& 3.215&3.896& -0.085& -0.538& -0.623 \\ 
 1024D&0.679 &3.193 &3.872& 0.694& 3.235&3.928& -0.085& -0.537& -0.622 \\ 
 4096S&0.692 &3.211 &3.902& 0.681& 3.214&3.895& -0.085& -0.537& -0.623 \\ 
 4096D&0.679 &3.193 &3.872& 0.693& 3.235&3.928& -0.085& -0.538& -0.623 \\
16384S&0.692 &3.211 &3.903& 0.681& 3.214&3.895& -0.085& -0.537& -0.623 \\ 
16384D&0.679 &3.193 &3.872& 0.694& 3.233&3.926& -0.085& -0.537& -0.622 \\
\hline
\end{tabular}

\hspace{0.5 cm}

The Sllod program used to generate the data in Table II reproduced earlier
Sllod results\cite{b57} for strain rates of 0.10 and 0.25 very well.  Gass'
Enskog-theory prediction\cite{b59} of the (linear) viscosity for the
conditions of Table II, $\eta _{\rm Enskog} = 1.5$, is just slightly higher
than the value $\eta = 1.25$ given by the data in this Table.  Just as in
the work of Refs. [10]-[12] there is no
indication of any logarithmic $N$-dependence in these results.  Let us now
compare these homogeneous results to those from boundary-driven simulations.

\subsection{Boundary-Driven Results with the Soft-Disk Potential}

With the soft-disk potential $\phi (r<1) = 100(1 - r^2)^4$ a useful
boundary-driven flow {\em does} result if the strain rate is moderate
and the system is not too large.  The boundary-driven flows {\em are}
sensitive to system size.  This is because the heat generated in a shear
flow, of order $L^D$ for $L$ large  eventually overwhelms the capacity
of the boundary, of order $L^{D-1}$, to absorb it.  The maximum
strain rate that can be reached by boundary-driven flows is therefore limited by
the heat conductivity as well as the efficiency of the frictional thermal
contact at the reservoir walls.  In addition to varying the size and stiffness of
the tethering potential, we also varied the lattice structure of the
tether sites, but settled on the simple square and cubic lattices when the
pressure-tensor results proved to be insensitive to lattice type.

Systems with Newtonian strips 20 atoms wide were already large enough that no
systematic deviation from linear stress behavior (with vanishing normal stress
difference, $\sigma _{xx} = \sigma_{yy} \simeq 0$), resulted.  Accordingly we
use a narrower system width here, 10, for an analysis of boundary driven
flows.  The system size is $N= 4 \times (10 \times 10) = 400$ particles.

Time-averaged boundary-driven velocity and normal-stress profiles are shown in the
figures.  The time-reversible frictional forces within the two
steadily-moving ``boundary'' chambers were chosen to maintain a kinetic
temperature in those chambers (relative to the tether velocity) of 0.70.
This approximately reproduces the conditions of the Sllod and Doll's
states in Table II.
Fig. 7 shows the average velocity
$\langle v_x(y) \rangle $, computed using the one-dimensional weight
function of Sec. VII.  For each of 400 equally-spaced gridpoints the
instantaneous values of the laboratory-frame velocity components
$\{ \dot x_i \}$ were spatially averaged at the grid points:
$$
v_x(y_G,t) \equiv
\sum _i \dot x_i w_{1D}(|y_G - y_i(t)|)/\sum _i w_{1D}(|y_G - y_i(t)|)
$$
These instantaneous gridpoint averages were then themselves averaged over
time and
are plotted in Fig. 7.

Pressure-tensor components at each particle were calculated in two
different ways.
Using the {\em assumed analytic form, a linear velocity profile},
$$
 0 < y < 10 \longrightarrow \langle v_x \rangle _{\rm analytic} = +5 \ ; \
$$
$$
10 < y < 20 \longrightarrow\langle v_x \rangle _{\rm analytic} = 15 - y \ ; \
$$
$$
20 < y < 30 \longrightarrow \langle v_x \rangle _{\rm analytic} = -5 \ ; \
$$
$$
30 < y < 40 \longrightarrow \langle v_x \rangle _{\rm analytic} = y - 35 \ ,
$$
the pressure tensors for
particles in the vicinity of each grid point were computed as averages,
using the two-dimensional weight function $w_{2D}(r<h=3)$.  These individual
pressure-tensor components were then averaged, with $w_{1D}(r<h=3)$ as
were the velocities of Fig. 7.  Pressure-tensor components were also computed
and averaged using the smooth-particle weights $w_{2D}(r<3)$ to calculate the
{\em instantaneous} flow velocity $\langle v(r_j)\rangle$ at each particle
$j$ rather than using the assumed linear profile.  The differences are
relatively small, as can be seen in Fig. 8, where the two approaches are
compared.  Using the instantaneous velocity at each particle is 
analogous to, but smoother than, the ``unbiased'' procedure
discussed by Evans and Morriss\cite{b16}.

The normal stress differences following either approach are considerably
larger than the Sllod results (Table II), and have the opposite sign to
the Doll's homogeneous results (Table II).
$$
\frac{P_{xx} - P_{yy}}{2} \simeq 0.03 \ ({\rm Boundary}{\rm -}{\rm Driven}) \ .
$$

The Sllod algorithm predicts a smaller effect (smaller by an order of
magnitude) while the Doll's algorithm
predicts the wrong sign!  We conclude that the two homogeneous algorithms
provide no more than an order of magnitude estimate of the normal stress
effects and further that these effects can be otherwise measured reliably, but with
some difficulty.

\begin{figure}
\includegraphics[height=6cm,width=6cm,angle=-0]{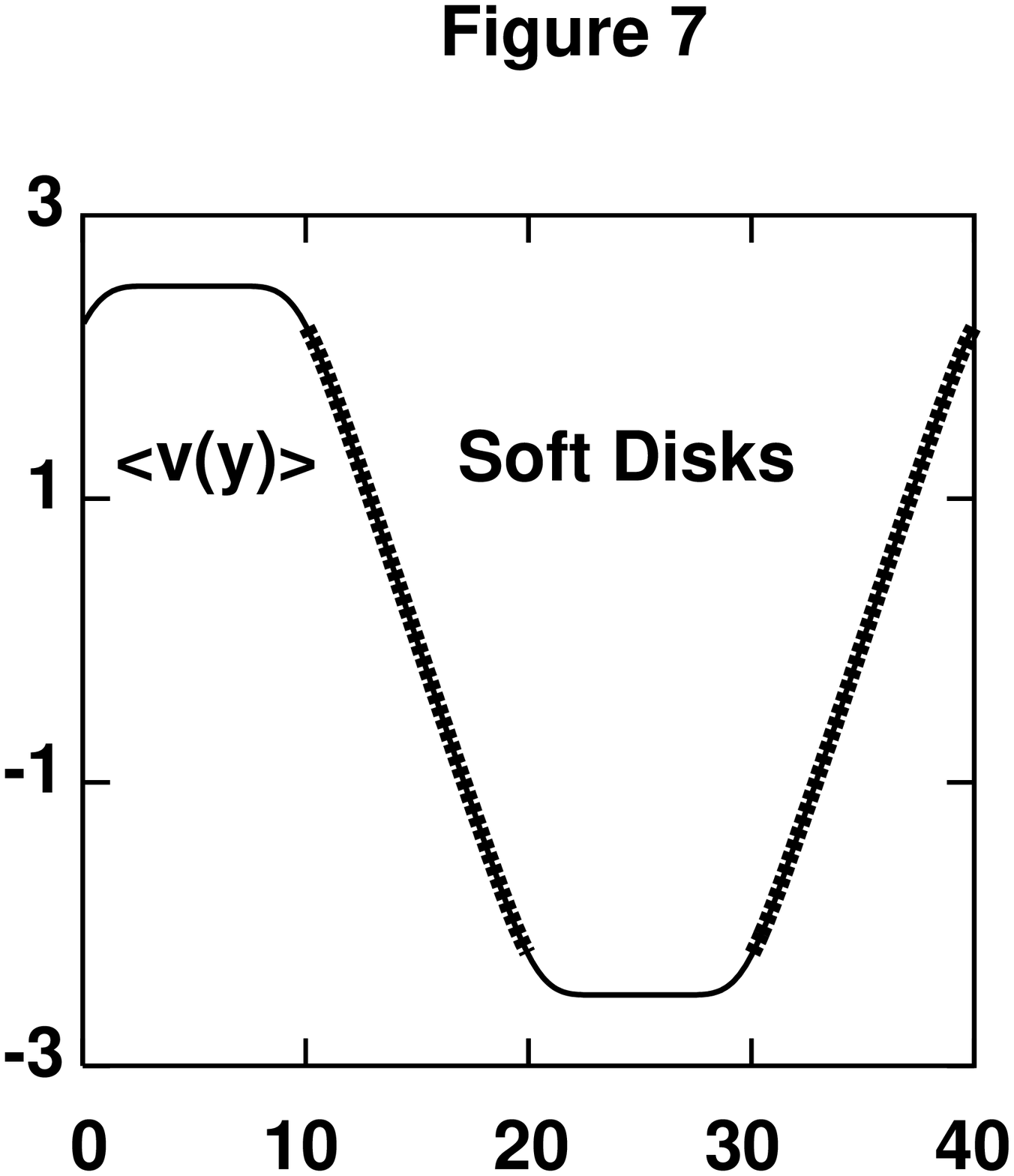}
\caption{
Velocity profile for a 400-particle system with the aspect ratio illustrated
in Fig. 4.  The tethering potential's force constant, $\kappa = 100$ 
provides an efficient coupling between the {\em driving} chambers at
$0 < y < 10$ and $20 < y < 30$ and the driven {\em Newtonian} chambers
at $10 < y < 20$ and $30 < y < 40$.  The locations of the two Newtonian
chambers are emphasized in the velocity plot.  The averaged velocity
profile was calculated with the smooth-particle weighting function
$w_{1D}(r<3)$.  The measured strain rate in the Newtonian regions is
about $\pm 0.48$.
}
\end{figure}

Fig. 8 shows stresses for the two portions of the four-chamber system
with $13 < y <17$ and $33 < y <37$.  These portions have ``typical''
bulk fluid averages, without any influence from the two driving
boundary regions.  This is a consequence of the smooth-particle
weight functions' range, $h=3$.

\begin{figure}
\includegraphics[bb=4.8in 2.1in 8.0in 5.0in,hiresbb=true,angle=-90]{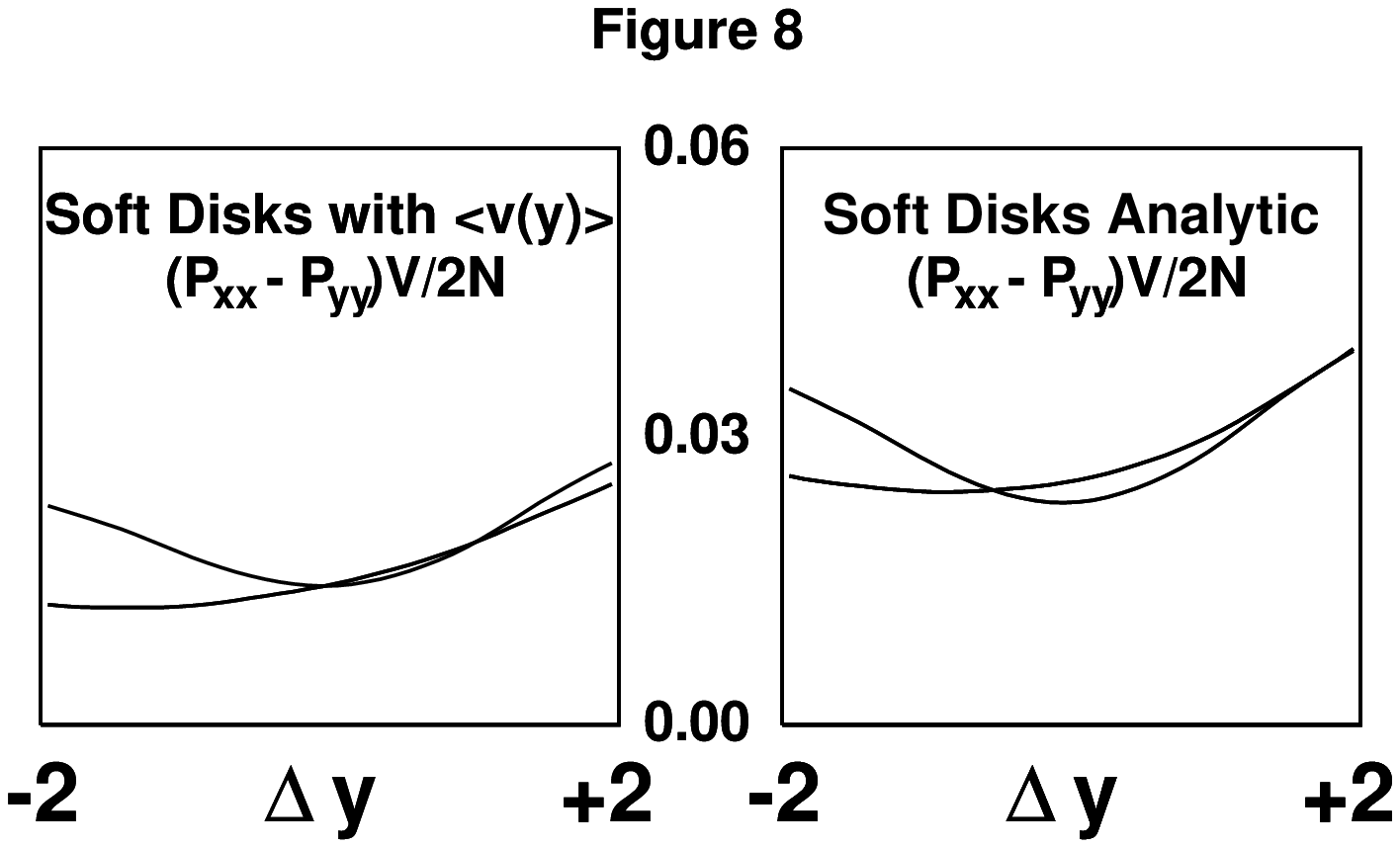}
\caption{
Boundary-driven two-dimensional flow using the ``soft-disk'' potential.
Average normal stress differences are shown.  At the left the stresses are
calculated relative to the instantaneous velocity profile.  There the
spatially-averaged instantaneous velocity and stress at each particle
are computed with $w_{2D}$, then averaged to get instantaneous
profiles using $w_{1D}$, and finally time averaged.  At the right the
stresses are calculated relative to an assumed linear velocity profile.
In both cases the stress difference is shown for the two
regions $13 < y < 17$ and $33 < y < 37$ free of boundary 
influences and hence typical of bulk fluid.  The run length was 5000.
}
\end{figure}

\begin{figure}
\includegraphics[height=6cm,width=6cm,angle=-0]{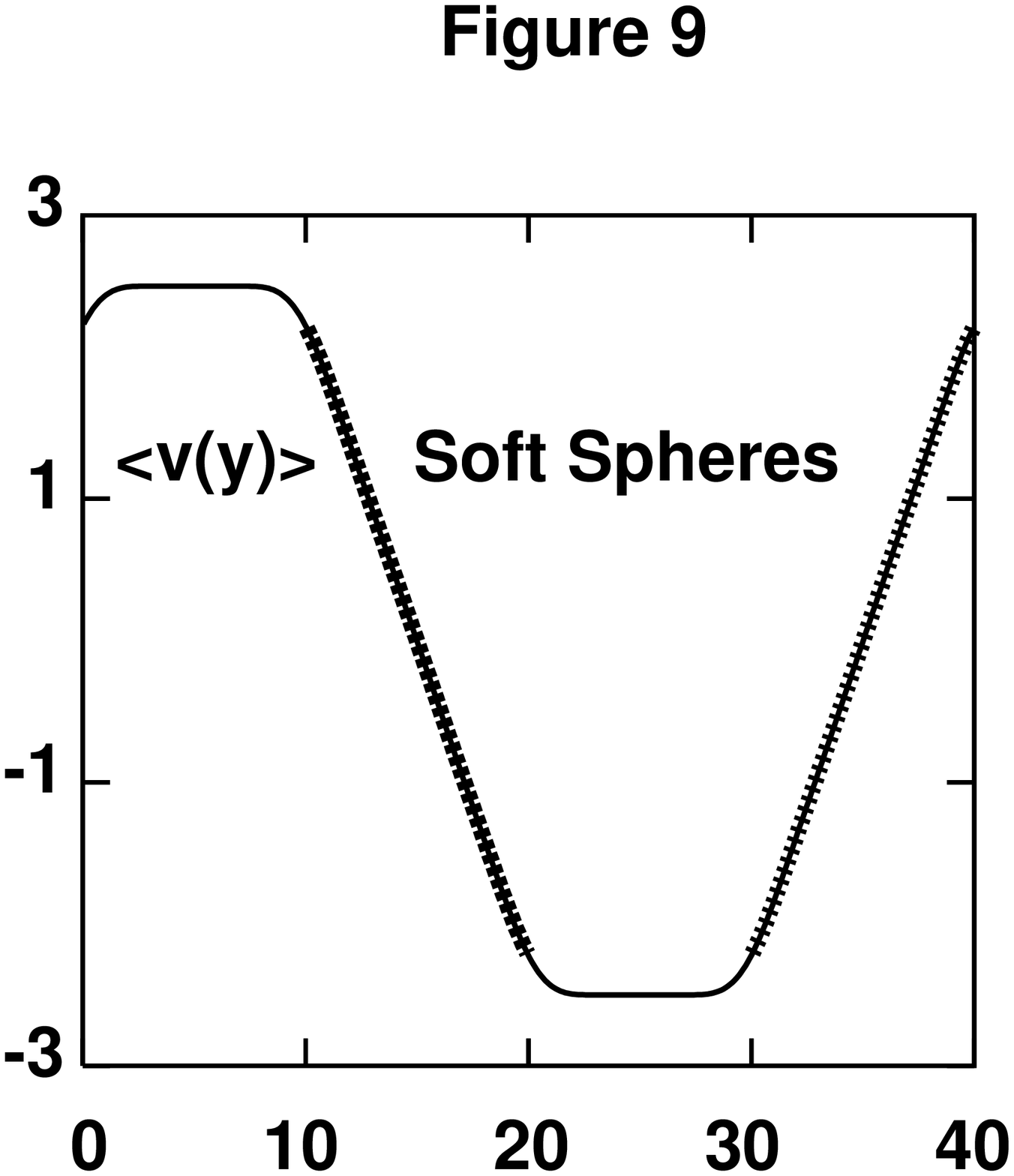}
\caption{
Time-averaged velocity profile for a 4000-particle three-dimensional
system with the aspect ratio illustrated
in Fig. 4.  The tethering potential's force constant, $\kappa = 100$ 
provides an efficient coupling between the {\em driving} chambers at
$0 < y < 10$ and $20 < y < 30$ and the driven {\em Newtonian} chambers
at $10 < y < 20$ and $30 < y < 40$.  The locations of the two Newtonian
chambers are emphasized in the velocity plot.  The averaged velocity
profile was calculated with the smooth-particle weighting function
$w_{1D}(r<3)$.  The measured strain rate in the Newtonian regions is
about $\pm 0.47$.  The run length was 1000.
}
\end{figure}

\section{Numerical Results: Three Dimensions}

\subsection{Periodic Shear}

Three-dimensional boundary-driven simulations require only the addition
of $z$ coordinates, with periodic boundary conditions in the $z$ direction.
For comparison purposes, we first generated series of isoenergetic Sllod
and Doll's periodic shears.  These results, shown in Table III, are for
periodic shearing of $L \times L \times L$ {\em cubes} of soft-{\em sphere}
fluid at unit density and energy:
$$
\phi = 100(1 - r^2)^4 \ ; \ r^2 = x^2 + y^2 + z^2 \ ;
\ N/V = Nm/V = E = K + \Phi \equiv 1 \ .
$$
The constant-energy ergostat forces keep the total energy of the
$N = L\times L \times L$ particles fixed.  The kinetic
energy $K$ is a sum in which each particle's contribution is measured
relative to the local velocity.  Choosing the cube center as the
coordinate origin, the systematic velocity in the $x$ direction is
taken to be proportional to $y$:
$$
\{ \ p^2/2m \equiv [p_x^2 + p_y^2 + p_z^2]/2m \ ; \
(p_x/m) \equiv v_x - \dot \epsilon y \ \} \ .
$$
The pressure-tensor results, given in Table III, are very insensitive
to system size.  Notice that the average kinetic temperature for all
these isoenergetic simulations is approximately 0.5.  The Sllod algorithm
gives
$$
T_{xx} > T_{yy} > T_{zz} \ {\rm (Sllod)} \ ,
$$
while the Doll's-Tensor algorithm gives instead
$$
T_{yy} > T_{xx} > T_{zz} \ {\rm (Doll's)} \ .
$$
The (correct) ``Boundary-Driven'' results, described next, show instead
the ordering
$$
T_{xx} > T_{zz} > T_{yy} \ {\rm (Boundary-Driven)} \ .
$$
The boundary-driven results also show qualitative differences from the
Sllod and Doll's results in the normal-stress differences, $(P_{xx} -
P_{yy})$ and $(P_{xx} - P_{zz})$.

\subsection{Boundary-Driven Shear}

We next implemented $4000 = 4 \times (10 \times 10 \times 10)$-particle
boundary-driven shear flows similar to the two-dimensional flows of Sec.
VIIB, but with periodic boundaries in the $z$ direction and with a fixed
isothermal boundary temperature 0.5, chosen to match the homogeneous periodic
results.  For comparison with these Sllod and Doll's results the same
nominal strain rate, $dv_x/dy \simeq 0.5$ was used.
Thus the two thermostated chambers move with velocities $(v_x = \pm 2.5)$.

The time-and-spatially-averaged velocity profile computed with the weight
function
$w_{1D}(|\delta y|<3)$ is shown in Figure 9.  The measured strain rates in the
straight-line portions of the profile are about $\pm 0.47$.  The energy
dissipation rate for 6000 thermostated degrees of freedom in the two moving
1000-particle reservoirs was
$6000 kT \langle \zeta \rangle = 6000 \times 0.5 \times 0.112 = 336$,
giving an estimate for the viscosity:
$$
\eta = T\dot S_{\rm external }/(V\dot \epsilon ^2) =
336/(2000 \times 0.5^2) = 0.672 \ ,
$$
within two percent of the periodic result, $0.688$ from Table III.  The
Newtonian shear stresses from the
$4000 = 4 \times (10 \times 10 \times 10)$-particle
simulation are $\pm 0.37$ in the bulk Newtonian regions, corresponding to
$$
\eta = -P_{xy}/(dv_x/dy) = 0.74 \ .
$$
The time-averaged spatially-smoothed normal stress differences,
$$
(P_{xx} - P_{yy})/2 \simeq 0.01 \ ; \ (P_{xx} - P_{zz})/2 \simeq 0.01 \ ,
$$
are only different with marginal significance, and
are shown in Figs. 10 and ll.  We considered three different system sizes,
to ensure that these results are insensitive to small geometrical changes.
Data for $N=4\times 10^3$ (run length 1000), $N=4\times 12^3$
(two runs of length 500; results from only one of them are shown here as
the difference between the two was insignificant), and $N=4\times 14^3$
(run length 400) are included in the figures.  Just as before,
the Newtonian regions for which the data are plotted are those free of
any boundary influences in the stress averaging.  The corresponding Sllod
and Doll's values for the stress differences, +0.003 and -0.016,
respectively, are quite different, just as in two dimensions.  The
disparity shows that neither homogeneous algorithm is even close to
``correct''.  The
actual difference between $P_{yy}$ and $P_{zz}$ is apparently quite small,
while both the Doll's and the Sllod algorithms indicate a relatively large
difference of order $\pm 0.04$.  The statistical fluctuations in the
boundary-driven simulations are not quite so large as to mask the
ordering of the two normal stress differences,
$$
P_{xx}-P_{yy} > P_{xx}-P_{zz} \ .
$$ 
Somewhat faster/larger computers could make this conclusion more
convincing.

\hspace{0.5 cm}

{\bf Table III}.  Soft-sphere viscosities in three dimensions with periodic
boundary conditions.
Space-and-time-averaged pressure tensors for homogeneous
Sllod and Doll's algorithms using the soft-sphere pair potential
illustrated in Fig. 6, $\phi = 100(1-r^2)^4$. The energy and density
are equal to unity and the strain rate,
$\dot \epsilon = dv_x/dy$ is 0.50.  The pressure-tensor components,
are given in the order $xx,yy,zz,xy$ with the kinetic, potential, and
total terms indicated.  The boundary conditions are periodic, with a
total run time of 200x10,000 timesteps for $N = 10 \times 10 \times
10 = 1000$ with a fourth-order Runge-Kutta timestep of 0.005.

\hspace{0.5 cm}

\begin{tabular}{| c || ccc || ccc || ccc || ccc |}
        \hline
$N $ &            $P_{xx}^K$ &$ P_{xx}^\Phi$ & $P_{xx}^\Sigma$ &
                  $P_{yy}^K$ &$ P_{yy}^\Phi$ & $P_{yy}^\Sigma$ &
                  $P_{zz}^K$ &$ P_{zz}^\Phi$ & $P_{zz}^\Sigma$ &
                  $P_{xy}^K$ &$ P_{xy}^\Phi$ & $P_{xy}^\Sigma$ \\
\hline
 216S&0.506 &2.012 &2.518& 0.497& 2.014&2.511& 0.493& 1.991&2.483 &
-0.062& -0.282& -0.344 \\
 216D&0.496 &2.004 &2.499& 0.507& 2.022&2.529& 0.493& 1.991&2.484 &
-0.062& -0.281& -0.342 \\
 512S&0.506 &2.010 &2.516& 0.497& 2.012&2.509& 0.493& 1.989&2.483 &
-0.063& -0.280& -0.343 \\
 512D&0.496 &2.001 &2.496& 0.507& 2.021&2.528& 0.493& 1.990&2.483 &
-0.063& -0.281& -0.343 \\
1000S&0.507 &2.009 &2.516& 0.497& 2.012&2.508& 0.493& 1.989&2.482 &
-0.063& -0.281& -0.344 \\
1000D&0.496 &2.000 &2.496& 0.507& 2.021&2.528& 0.493& 1.989&2.482 &
-0.063& -0.280& -0.343 \\
1728S&0.507 &2.009 &2.516& 0.497& 2.012&2.509& 0.493& 1.988&2.481 &
-0.063& -0.281& -0.344 \\
1728D&0.496 &2.001 &2.497& 0.508& 2.020&2.528& 0.493& 1.989&2.481 &
-0.063& -0.281& -0.344 \\

2744S&0.507 &2.009 &2.516& 0.497& 2.012&2.509& 0.493& 1.989&2.481 &
-0.063& -0.281& -0.344 \\
2744D&0.496 &2.000 &2.496& 0.508& 2.020&2.528& 0.493& 1.989&2.482 &
-0.063& -0.280& -0.343 \\

\hline
\end{tabular}

\hspace{0.5 cm}

\begin{figure}
\includegraphics[bb=6.0in 0.0in 8.0in 5.5in,hiresbb=true,angle=-90]{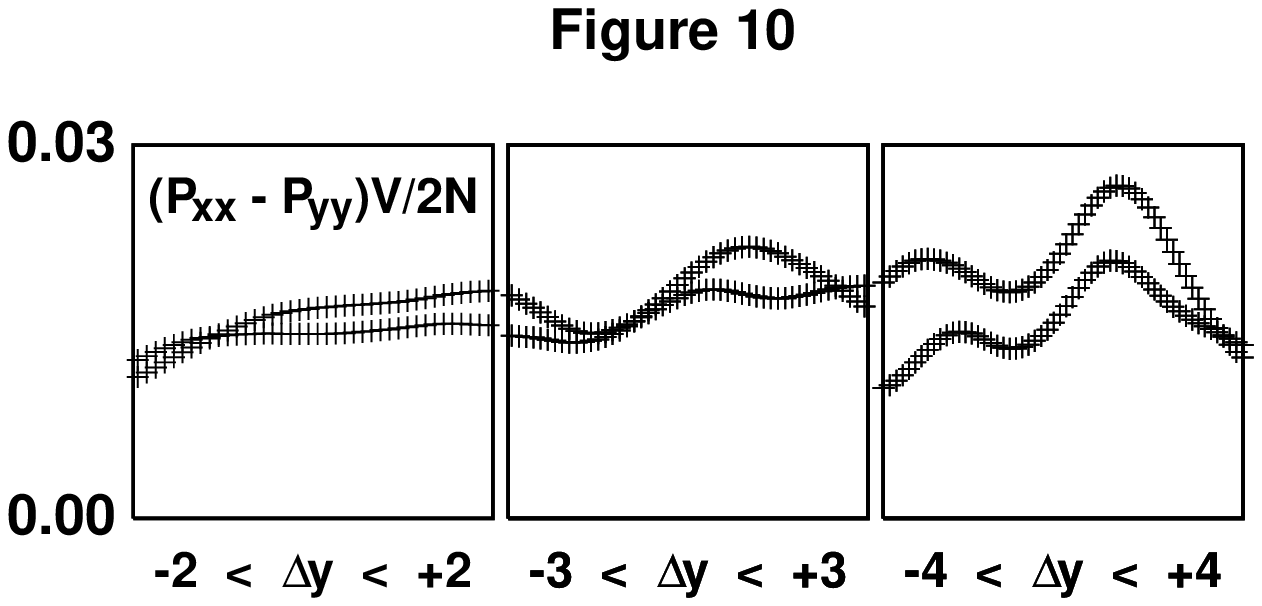}
\caption{
Normal stress differences for boundary-driven three-dimensional flows using
the ``soft-sphere'' potential.  The normal stress difference
$(P_{xx} - P_{yy})V/2N$ is shown here for three different system sizes,
with stress calculated relative to the {\em instantaneous} velocity
profile. The spatially-averaged instantaneous particle values of
velocity and stress are computed with $w_{3D}$.  Then the particle
values are averaged to get instantaneous profiles using $w_{1D}$.
The figure shows time averages of those instantaneous profiles.  Only data
from the two distinct regions free of boundary averaging influences are
shown here. 
}
\end{figure}

As the system size is increased, with the strainrate fixed, the
Newtonian temperature increases also, in rough accord with the linear
model treatment of Sec. II.  That is, the central temperature increase
is proportional to $L^2$.  Figure 12 shows temperature profiles for
the three system sizes considered here.  We consider the kinetic temperature
here, because of its relative conceptual simplicity and its
physical conceptual basis\cite{b37}.  In the boundary-driven shear flows
the {\em ordering} of the kinetic temperatures is
$$
T_{xx} > T_{zz} > T_{yy} \ ,
$$
with the difference between $T_{xx}$ and $ T_{zz}$ two or three times larger
than that between $T_{zz}$ and $T_{yy}$.
\begin{figure}
\includegraphics[bb=6.0in 0.0in 8.0in 5.5in,hiresbb=true,angle=-90]{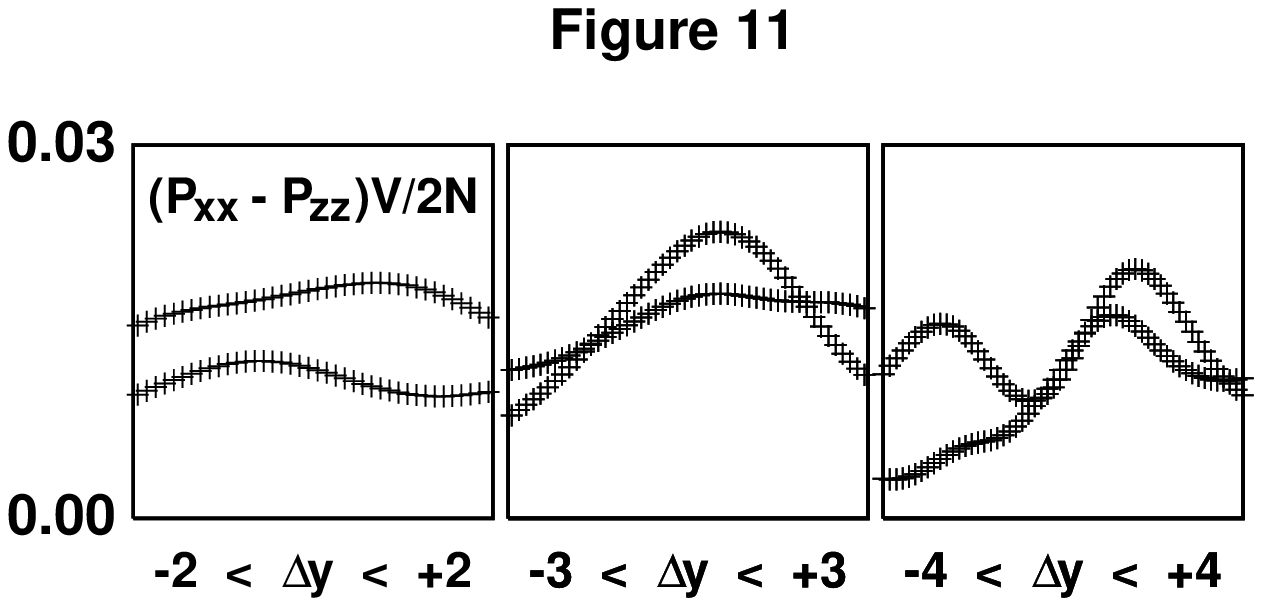}
\caption{
Normal stress differences for boundary-driven three-dimensional flows using
the ``soft-sphere'' potential.  The normal stress difference
$(P_{xx} - P_{zz})V/2N$ is shown here for three different system sizes,
with stress calculated relative to the {\em instantaneous} velocity
profile. The spatially-averaged instantaneous particle values of
velocity and stress are computed with $w_{3D}$.  Then the particle
values are averaged to get instantaneous profiles using $w_{1D}$.
The figure shows time averages of those instantaneous profiles.  Only data
from the two distinct regions free of boundary averaging influences are
shown here.
}
\end{figure}

\begin{figure}
\includegraphics[bb=5.1in 0.0in 8.0in 5.5in,hiresbb=true,angle=-90]{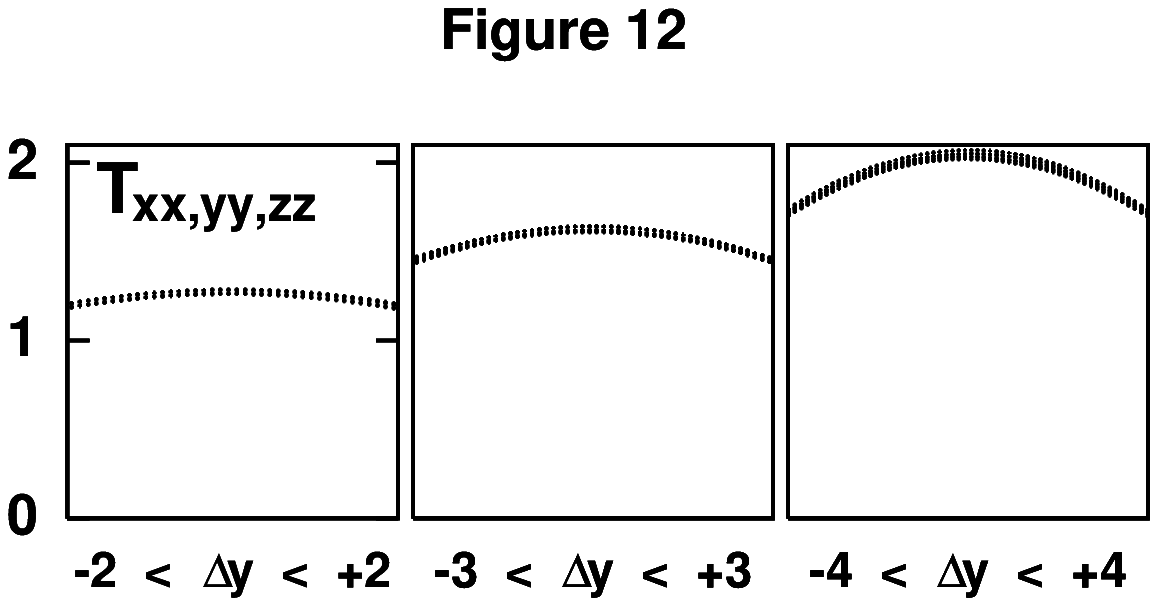}
\caption{
Tensor temperatures for boundary-driven three-dimensional flows using
the ``soft-sphere'' potential.  The time-averaged kinetic temperatures,
relative to the instantaneous velocity profile, computed with $w_{3D}$ and
$w_{1D}$, are shown here for three system sizes, all with boundary
temperatures and strain rates equal to 0.5.  The systems contain
four chambers with $10^3$ (run length 1000), $12^3$ (run length 500), and
$14^3$ (run length 400) particles per chamber in the three cases shown.
Although data from the two distinct regions free of boundary averaging
influences are shown here the differences between $T_{xx}$, $T_{yy}$,
and $T_{zz}$, are of order $\pm 0.02$ and are too small to see on the
scale of the figure.
}
\end{figure}

\section{Summary and Conclusions}

We were able to characterize the tensor temperature and the nonlinear
stresses for both homogeneous and boundary-driven versions of simple shear.
Neither Sllod nor Doll's gives the correct ordering of the kinetic
temperatures $\{ \ T_{ii} \ \}$.  Generally the ``Sllod'' algorithm gives
a somewhat ``better'' approximation to the normal-stress differences,
$\{ \ P_{ii} - P_{jj} \ \}$, though Sllod is certainly far from ``correct''.
Despite its evident failures, there is a fairly widespread faith in the Sllod
approach\cite{b60}.  It is clear (as emphasized to us by Jim Lutsko; see also
ref. [61]) that the
algorithms' extra rotational terms in the motion equations,
$-\dot \epsilon p_x$ for Sllod and $-\dot \epsilon p_y$ for Doll's, when
left to their own devices, would eventually cause
$p_x^2$ and $T_{xx}$ to diverge for the Sllod algorithm, and $p_y^2$ and
$T_{yy}$ to diverge for Doll's.  This provides a clear explanation of the
qualitative difference between the two algorithms' predictions and the
corresponding opposite directions for the rotation of the principal axis
of the stress.

Lutsko\cite{b62} has reviewed the hard-sphere-based Enskog theory for nonlinear
stress (``the only viable theory'') and his finding that $P_{xx} > P_{yy}$
in simple shear is quite consistent with our results.  On the other hand
some theoretical models\cite{b63} and some computer simulations\cite{b16}
find $P_{xx} < P_{yy}$, even for relatively simple fluids, so it is clear
that more investigations are required.  Evidently the temperature tensor
and the nonlinear stresses are {\em not} given accurately by the Sllod
algorithm.  The more realistic boundary-driven flows need to be used whenever
confidence in the results is required.

Boundary-driven flows are actually extremely complex, even for this
simplest possible model of shear.  The flows we can study are dominated
by fluctuations which can be tamed by averaging, in one, two, or three
dimensions, but the time-averaged flows describe the time-dependent situation
no better than they would for a physical waterfall or a turbulent stream.

It is fortunate that a hydrodynamic description of flows {\em is} feasible on a
very small scale (just a few particle diameters) as was apparent from the
earliest shockwave simulations, which showed shockwidths of only a few
particle diameters. It still remains a puzzle that shockwaves indicate an
enhanced nonlinear viscosity while the homogeneous shear algorithms
considered here predict a reduction rather than an enhancement\cite{b45}.

\section{Acknowledgments}

We thank Karl Travis and Billy Todd for providing some relevant
literature and for stimulating our interest in the
differences between the algorithms.  Denis Evans kindly stated that there
was no compelling reason to favor the Sllod algorithm for stationary flows.
Debra Bernhardt made several useful comments on the manuscript.
Jim Lutsko provided useful references and pointed out the physical reason
underlying the positive nature of the normal stress difference
$P_{xx} - P_{yy} $.  The same explanation accounts also for the positive
sign of $P_{xx} - P_{zz} $.  See also the low-density Boltzmann equation
treatment of shear flow\cite{b61}.  We had considerable correspondence
seeking evidence for logarithmic divergence of the two-dimensional shear
viscosity, but found none.  We specially thank Arek Bra\'nka\cite{b14}, Giovanni
Ciccotti\cite{b15}, Michio Otsuke, and Hiroshi Watanabe for references,
comments, and suggestions.


\begin{thebibliography}{99}


\bibitem{b1}  Wm. G. Hoover, {\em Computational Statistical Mechanics}
              (Elsevier, Amsterdam, 1991, available at the homepage
              http://williamhoover.info/book.pdf).

\bibitem{b2}  J. Petravic, J. Chem. Phys. {\bf 127}, 204702 (2007).

\bibitem{b3}  Wm. G. Hoover, {\em Smooth Particle Applied Mechanics ---
              The State of the Art} (World Scientific Publishers, Singapore,
              2006, available from the publisher at the publisher's site
              http://www.worldscibooks.com/mathematics/6218.html).

\bibitem{b4}  W. G. Hoover, Lecture Notes in Physics {\bf 132}, 373
              (Springer-Verlag, Berlin, 1985).

\bibitem{b5}  D. J. Evans and G. P. Morriss, Phys, Rev. A {\bf 30}, 
              1528 (1984).

\bibitem{b6}  Wm. G. Hoover and W. T. Ashurst, {\em Theoretical
              Chemistry, Advances and Perspectives}
               {\bf 1}, 1 (Academic, New York, 1975).

\bibitem{b7}  W. T. Ashurst, Ph. D. thesis, University of California
              at Davis, 1974.

\bibitem{b8}  W. T. Ashurst and W. G. Hoover, Phys. Rev. Lett.
              {\bf 31}, 206 (1973).

\bibitem{b9}  S. Y. Liem, D. Brown and J. H. R. Clarke, Phys. Rev. A,
              {\bf 45}, 3706 (1992).

\bibitem{b10} W. G. Hoover, W. T. Ashurst, and R. J. Olness, J. Chem.
               Phys. {\bf 60}, 4043 (1974).

\bibitem{b11} T. Keyes and I. Oppenheium, Phys. Rev. A {\bf 8},
              937 (1973).

\bibitem{b12} D. Gravina, G. P. F. Ciccotti, and B. L. Holian, Phys.
              Rev. E {\bf 52}, 6123 (1995).

\bibitem{b13} B. Liu and J. Goree, Phys. Rev. Lett. {\bf 94}, 185002 (2005).

\bibitem{b14} A. C. Bra\'nka and D. M. Heyes, Phys. Rev. E {\bf 55}, 5713
              (1997).

\bibitem{b15} M. Ferrario, A. Fiorino, and G. P. F. Ciccotti, Physica A
              {\bf 240}, 268 (1997).

\bibitem{b16} D. J. Evans and G. P. Morriss, {\em Statistical Mechanics of
              Nonequilibrium Liquids} (Academic, London, 1990, available
              at the authors' webpages.

\bibitem{b17} J. L. Tuck and M. T. Menzel, Adv. Math. {\bf 9}, 399 (1972).

\bibitem{b18} B. J. Alder and T. E. Wainwright, J. Chem. Phys.
              {\bf 31}, 459 (1959).

\bibitem{b19} J. B. Gibson, A. N. Goland, M. Milgram, and G. H. Vineyard,
              Phys. Rev. {\bf 120}, 1229 (1960).

\bibitem{b20} J. A. Barker and D. Henderson, Annual Review of Physical
              Chemistry {\bf 23}, 439 (1972).

\bibitem{b21} J. F. Lutsko, Phys. Rev. Lett. {\bf 78}, 243 (1997).

\bibitem{b22} Wm. G. Hoover, K. Boercker, and H. A. Posch, Phys. Rev. E
              {\bf 57}, 3911 (1998).

\bibitem{b23} W. G. Hoover, D. J. Evans, R. B. Hickman, A. J. C. Ladd,
              W. T. Ashurst, and B. Moran, Phys. Rev. A {\bf 22}, 1690 (1980).

\bibitem{b24} M. J. Gillan and M. Dixon, J. Phys. C {\bf 16}, 869 (1983).

\bibitem{b25} D. J. Evans, Phys. Lett. A {\bf 91}, 457 (1982).

\bibitem{b26} R. Zwanzig, Annual Review of Physical Chemistry {\bf 16},
              67 (1965).

\bibitem{b27} K. Kadau, T. C. Germann, and P. S. Lomdahl,  Int. J.
              Mod. Phys. C {\bf 17}, 1755 (2006).

\bibitem{b28} M. Mareschal, J-P. Ryckaert, and A. Bellemans, Mol. Phys.
              {\bf 61}, 33 (1987).

\bibitem{b29} P. J. Daivis, M. L. Matin, and B. D. Todd, J.
              Non-Newtonian Fluid Mech. {\bf 111}, 1 (2003).

\bibitem{b30} B. D. Todd and P. J. Daivis, Phys. Rev. Lett. {\bf 81}, 1118
              (1998).

\bibitem{b31} A. Baranyai and P. T. Cummings, J. Chem. Phys. {\bf 110}, 42
              (1999).

\bibitem{b32} B. D. Todd and P. J. Daivis, Comp. Phys. Comm. {\bf 117}, 191
              (1999).

\bibitem{b33} D. M. Heyes, Chem. Phys. {\bf 98}, 15 (1985).

\bibitem{b34} P. J. Daivis and B. D. Todd, J. Chem. Phys. {\bf 124}, 194103 (2006).

\bibitem{b35} B. J. Edwards, C. Baig, and D. J. Keffer, J. Chem. Phys.
              {\bf 124}, 194104 (2006).

\bibitem{b36} B. D. Todd and P. J. Daivis, Mol. Sim. {\bf 33}, 189 (2007).

\bibitem{b37} Wm. G. Hoover and C. G. Hoover, Phys. Rev. E 77, 041104 (2008).

\bibitem{b38} D. J. Evans, E. G. D. Cohen, and G. P. Morriss, Phys.
              Rev. Lett. {\bf 71}, 2401 (1993).

\bibitem{b39} D. J. Evans and D. J. Searles, Phys. Rev. E {\bf 50}, 1645 (1994).

\bibitem{b40} G. Gallavotti and E. G. D. Cohen, Phys. Rev. Lett. {\bf 74},
              2694 (1995).

\bibitem{b41} B. L. Holian, W. G. Hoover, and H. A. Posch, Phys. Rev. Lett.
              {\bf 59}, 10 (1987).

\bibitem{b42} S. Nos\'e, J. Chem. Phys. {\bf 81}, 511 (1984).
 
\bibitem{b43} W. G. Hoover, Phys. Rev. A {\bf 31}, 1695 (1985).
  
\bibitem{b44} W. G. Hoover, A. J. C. Ladd, and B. Moran, Phys. Rev. Lett.
              {\bf 48}, 1818 (1982).

\bibitem{b45} B. L. Holian, W. G. Hoover, B. Moran, and G. K. Straub,
              Phys. Rev. A {\bf 22}, 2798 (1980).

\bibitem{b46} Wm. G. Hoover, Physica A {\bf 240}, 1 (1997).

\bibitem{b47} C. P. Dettmann and G. P. Morriss, Phys. Rev. E {\bf 54}, 2495 (1996).

\bibitem{b48} A. W. Lees and S. F. Edwards, J. Phys. C {\bf 5}, 1921 (1972).

\bibitem{b49} W. Prager, {\em Introduction to the Mechanics of Continua}
              (Ginn \& Company, Boston, 1961 [Dover reprint, 1973]).

\bibitem{b50} Wm. G. Hoover and H. A. Posch, Phys. Rev. E {\bf 54}, 5142 (1996).

\bibitem{b51} Wm. G. Hoover, K. Aoki, C. G. Hoover
              and S. V. De Groot, Physica D {\bf 187}, 253 (2004).

\bibitem{b52} K. Aoki, and D. Kusnezov, Phys. Lett. A {\bf 265}, 250 (2000).

\bibitem{b53} L. B. Lucy, The Astronomical Journal {\bf 82}, 1013 (1977).

\bibitem{b54} Wm. G. Hoover, C. G. Hoover, and E. C. Merritt, Phys. Rev. E
              {\bf 69}, 016702 (2004).

\bibitem{b55} O. Kum, Wm. G. Hoover, and H. A. Posch, Phys. Rev. E {\bf 52},
              4899 (1995).

\bibitem{b56} H. A. Posch, Wm. G. Hoover, and O. Kum, Phys. Rev. E
              {\bf 52}, 1711 (1995).

\bibitem{b57} Wm. G. Hoover and H. A. Posch, Phys. Rev. E {\bf 51},
              273 (1995).

\bibitem{b58} P. J. Daivis, J. Non-Newtonian Fluid Mech. {\bf 152}, 120
              (2008).

\bibitem{b59} D. M. Gass, J. Chem. Phys. {\bf 54}, 1898 (1971).

\bibitem{b60} Ch. Dellago and H. A. Posch, ``Shear Viscosity and Lyapunov
              Instability of a Hard-Disk Couette Flow'' (preprint, April 2008).

\bibitem{b61} A. J. C. Ladd and W. G. Hoover, J. Stat. Phys. {\bf 38},
              973 (1985).

\bibitem{b62} J. F. Lutsko, Phys. Rev. E {\bf 58}, 434 (1998).

\bibitem{b63} H. H. Gan and B. C. Eu, Phys. Rev. A {\bf 46}, 6344 (1992).


\end{thebibliography}
\end{document}